\pdfoutput=1
\documentclass{article} 
\usepackage{subfigure}
\usepackage{graphicx} 
\usepackage{pdfpages} 
\usepackage{amssymb}
\usepackage{amsmath}
\usepackage{url}
\usepackage{appendix}
\usepackage{booktabs}
\usepackage{comment}
\usepackage{float}
\usepackage{tikz}

\makeatletter
\def\@normalsize{\@setsize\normalsize{10pt}\xpt\@xpt
\abovedisplayskip 10pt plus2pt minus5pt\belowdisplayskip 
\abovedisplayskip \abovedisplayshortskip \z@ 
plus3pt\belowdisplayshortskip 6pt plus3pt 
minus3pt\let\@listi\@listI}

\def\subsize{\@setsize\subsize{12pt}\xipt\@xipt}
\def\section{\@startsection {section}{1}{\z@}{1.0ex plus
1ex minus .2ex}{.2ex plus .2ex}{\large\bf}}
\def\subsection{\@startsection 
   {subsection}{2}{\z@}{.2ex plus 1ex} {.2ex plus .2ex}{\subsize\bf}}
\makeatother
\begin{document}
\date{}

\title{\bf Linearity and correction of the BF effect in LSST sensors}

\author{Craig Lage\\
  Department of Physics\\
  University of California - Davis\\
cslage@ucdavis.edu}

\maketitle

\subsection*{\centering Abstract}

{\em Keywords: 
LSST, modeling, camera, CCD, simulation, diffusion, image processing.  
}
\\
The Brighter-Fatter (hereafter BF) effect in CCD sensors causes increases in the image size of bright objects due to electrostatic repulsion of collected charges.  Correcting this effect in the LSST camera is required in order to meet the science goals of the project, especially galaxy shape measurements for weak lensing.  The current plan for BF image correction in the LSST is to use the deconvolution method described in Coulton, et.al. \cite{Coulton_2018}.  In this work, we study the linearity of the BF effect and effectiveness of the Coulton correction, using both simulation tools and measurements made on prototype LSST CCDs from both CCD vendors.  We conclude that the proposed image correction method may be adequate to meet the LSST science goals, although more work is needed on the algorithms used to generate the image correction kernel from sensor measurements.

\section{Introduction}
The Large Synoptic Survey Telescope (LSST) is an innovative, large, fast survey telescope currently under construction at Cerro Pachon in Chile \cite{LSST_2019}.  The digital camera for the LSST, also currently under construction, will consist of approximately 3.2 gigapixels and will be the largest digital camera ever constructed.  The camera uses fully-depleted silicon Charge Coupled Devices (CCDs) which are back illuminated and 100 microns thick in order to optimize quantum efficiency in the near infrared.  The imaging area consists of 189 CCDs, with each CCD containing 16 imaging regions laid out in an 8x2 array.  Each imaging region has a pixel array with approximately 500x2000 10 micron square pixels, giving 16 Megapixels total.  Each imaging region also has its own independent amplifier (\cite{oconnor2019uniformity}, \cite{oconnor_2016}).  The LSST focal plane contains CCDs from two different vendors, the ITL STA3800C from the  University of Arizona Imaging Technology Laboratory \cite{ITL_website}, and the E2V CCD250 from Teledyne E2V \cite{E2V_website}.  

One of the major science goals of the LSST is to use weak lensing techniques to map the distribution of mass in the universe as a function of redshift, thereby gaining insight into the distribution of dark matter, and the nature and evolution of dark energy.   Since the LSST will dramatically increase the number of galaxies being mapped, statistical errors will be reduced.  Systematic errors will then become dominant, so reducing the systematic errors of galaxy shape measurements will be required in order to meet the LSST science goals.  Doing this requires detailed knowledge of the point spread function (PSF) of the telescope so that accurate measurements of galaxy shapes can be made on a large number of distant galaxies.  The PSF is typically modeled across the focal plane using measurements of Milky Way stars.  One factor impacting this PSF measurement is the Brighter-Fatter (BF) effect.  As is well known (see, for example \cite{antilogus2014}, \cite{guyonnet2015}, \cite{gruen2015}, \cite{Coulton_2018}), the BF effect causes the sizes of stellar images to increase as the brightness of the star increases.  This is due to charges stored in the CCD collecting wells electrostatically repelling additional incoming charges and driving them into surrounding pixels.  What is perhaps less well known is that, because of asymmetry in the construction of the CCD, the BF effect also results in increasing ellipticity as the star's brightness increases.  This is because the charge confinement in the X and Y directions of the CCD is fundamentally different, with charge being confined in one direction by the introduction of implanted channel stops, and with charge being confined in the other direction by the application of varying parallel gate voltages.  Thus, we find that it is important to correct both the size increase and the ellipticity increase in order to reduce the systematic errors in galaxy shape determination and meet the weak lensing goals of the LSST project.

The current plan for BF image correction in the LSST is to use the deconvolution method described in Coulton, et.al. \cite{Coulton_2018}.  Several recent studies (\cite{rasmussen2016}, \cite{astier2019}) have raised concerns about second-order effects which may limit the effectiveness of this proposed correction method.   To study this issue, in this work we use a combination of CCD simulation techniques and CCD measurements to investigate several possible sources of these second order effects and to evaluate how well the planned correction technique works for correcting images of artificial stars in the laboratory.

This paper is divided into several sections.  In the first section, we use simulation techniques to study departures from linearity of the pixel distortions as the stored charge in a pixel increases, as well as departures from the assumption that the distortions due to charges in nearby pixels can be linearly superposed.  The second section examines second order effects in the pixel-pixel covariances seen in flat images.   Then we review how successfully the current correction algorithms remove the BF effect on images of artificial stars, and discuss proposals to improve the algorithms which generate the BF correction kernel.  We then summarize the results and make recommendations for future work.

\section{Simulation study of BF non-linearity}
This section describes the charge transport simulation methods that are used to study the BF correction, and details the results obtained.  To perform the simulations described here, we have built a numerical Poisson solver which solves Poisson's equation on a 3D lattice which describes the silicon volume of the CCD.  The basic simulation methods are described in \cite{Lage_2017} and \cite{Poisson-CCD-paper}.  The simulation code itself, with many examples, is available at \cite{Poisson-CCD-code}.  CCDs from both vendors were submitted for physical analysis and measurements of the CCDs were obtained, including dopant profiles and physical dimension measurements of the silicon structures, in order to inform the simulations and remove most degrees of freedom \cite{CCD-Physical-Analysis}.   An important result of the simulation is the shape of the charge packet stored in the collecting wells.  The charge packet is in equilibrium between the electric field, which tends to push the charges toward the center of the well, and diffusion, which tends to push the charges apart.  In equilibrium, the charge packet shape is such that the net current flow, including both drift current and diffusion current, is zero.  This condition is met when the quasi-Fermi level is constant in each collecting well (see, for example \cite {sze1981physics}).  With this constraint the simulation is able to self-consistently solve for both the potentials in the silicon volume and the free carrier densities.  This enables studies of how the stored charge packet builds up during image integration, as well as study of how one charge packet influences the shape of adjacent charge packets.

\subsection{Simulation of the charge packet and pixel distortions}
\begin {figure}[p]
    \centering
	\subfigure[b][Charge packet with 100,000 electrons]{\includegraphics[trim=0.0in 0.0in 0.0in 0.0in,clip,width=0.40\textwidth]{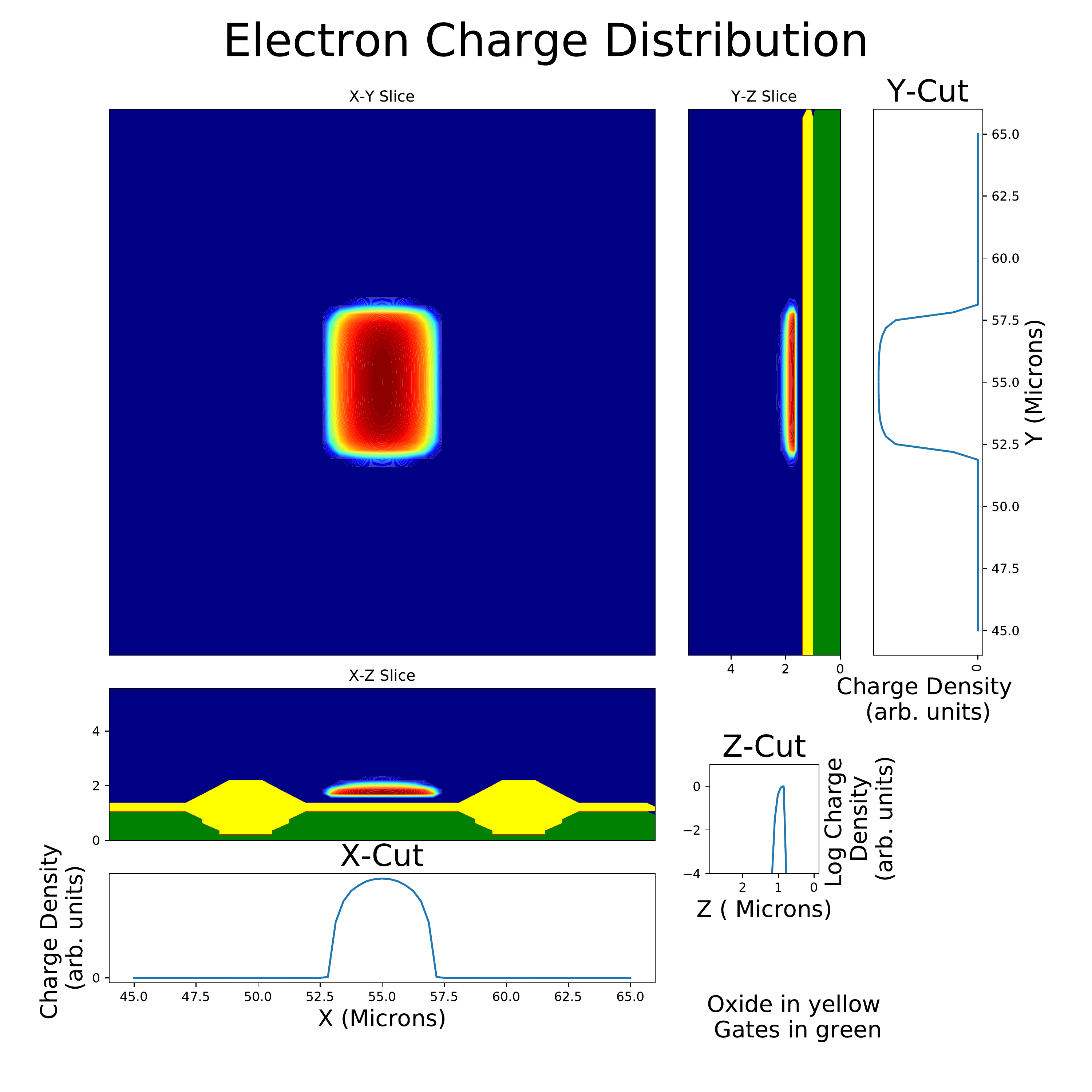}}    
	\subfigure[b][Pixel distortion from the central charge packet]{\includegraphics[trim=1.0in 0.0in 1.0in 0.0in,clip,width=0.60\textwidth]{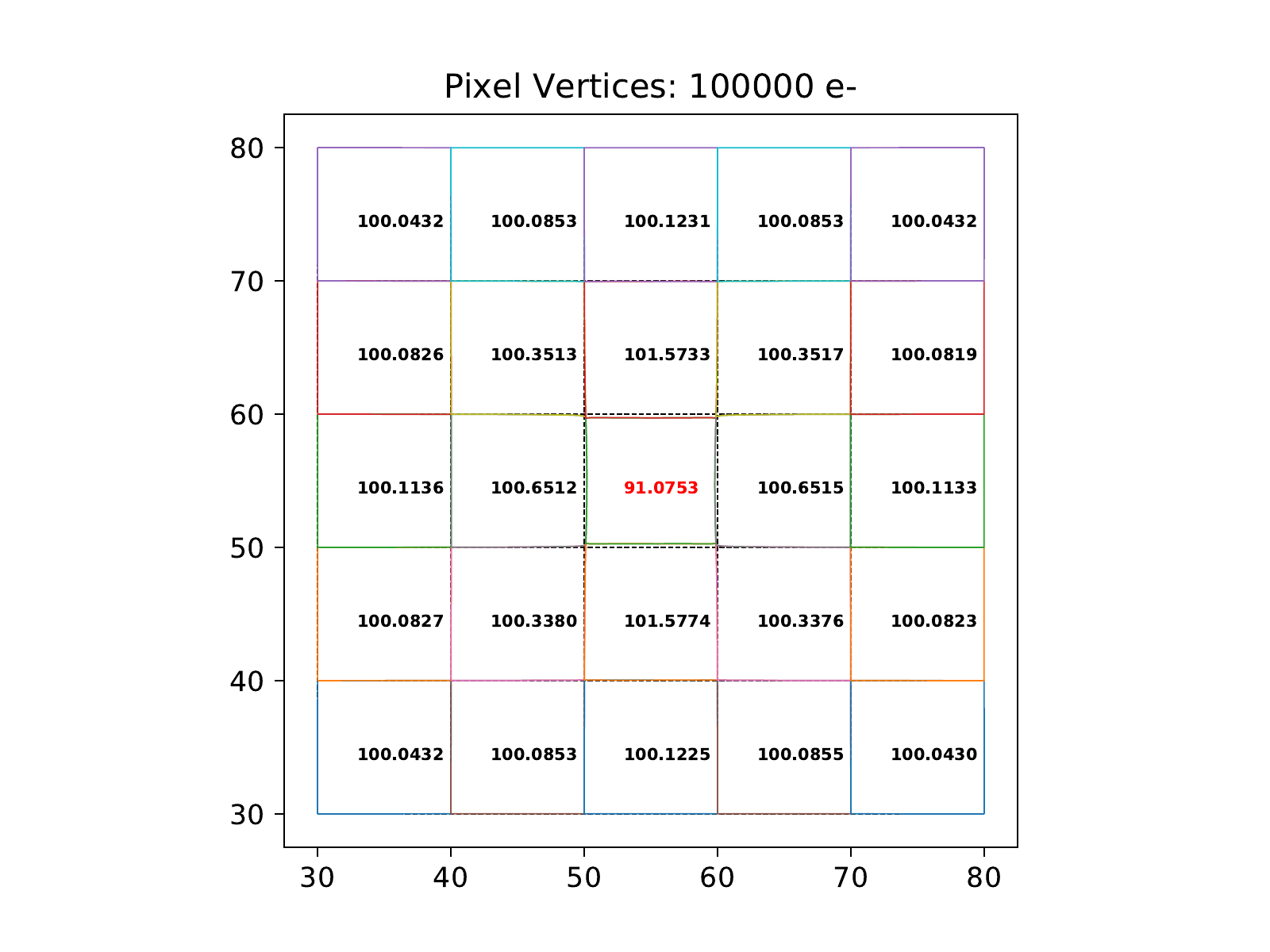}}
  \caption{Simulation of pixel distortions in an ITL chip when the central pixel contains 100,000 electrons and the surrounding pixels are empty.  X and Y are the lateral dimensions of the CCD, and Z is the thickness dimension.  The CCDs are 100 microns thick.  These distortions are obtained by solving Poisson's equation for the potentials in the CCD, then tracking electrons down and using a binary search to determine the pixel boundaries.  As expected, the central pixel loses area and the surrounding pixels all gain area. Note that the loss in area of the central pixel is greater than the sum of the area gains of the surrounding pixels because there are more distant pixels which are not plotted here and which also gain area.}
  \label{Pixel_Distortions}
  \end{figure}

In order to characterize the BF effect, we first simulate a situation where one pixel has a fixed amount of charge (typically 100,000 electrons), and all surrounding pixels are empty.  After solving for the potential and resulting electric field, we can track electrons down through the silicon.  As the electrons travel down through the silicon under the influence of the electric field, they eventually end up in one of the collecting wells.  A binary search is used to find the bifurcation points where electrons on one side of the bifurcation point end up in one pixel, and electrons on the other side of the bifurcation point end up in an adjacent pixel.   These are identifed as the pixel boundaries.  This allows us to characterize the distortion in the pixel boundaries which results from the central pixel charge.  A typical result of this process is shown in Figure \ref{Pixel_Distortions}.  The distorted pixel shapes which result are what gives rise to the BF effect, with the central pixel losing area, which is gained by the surrounding pixels.  As we will see in the subsequent sections, these changes in the pixel shapes can be measured using covariances extracted from flat images.  References \cite{antilogus2014} and \cite{Coulton_2018} give more detail on the relation between the pixel distortions and the pixel-pixel covariances seen in flat images.

\subsection{Charge packet build-up and linearity of the pixel distortions}
\begin {figure}[p]
	\centering
	\subfigure[b][10K electrons]{\includegraphics[trim=0.0in 0.0in 0.0in 0.0in,clip,width=0.45\textwidth]{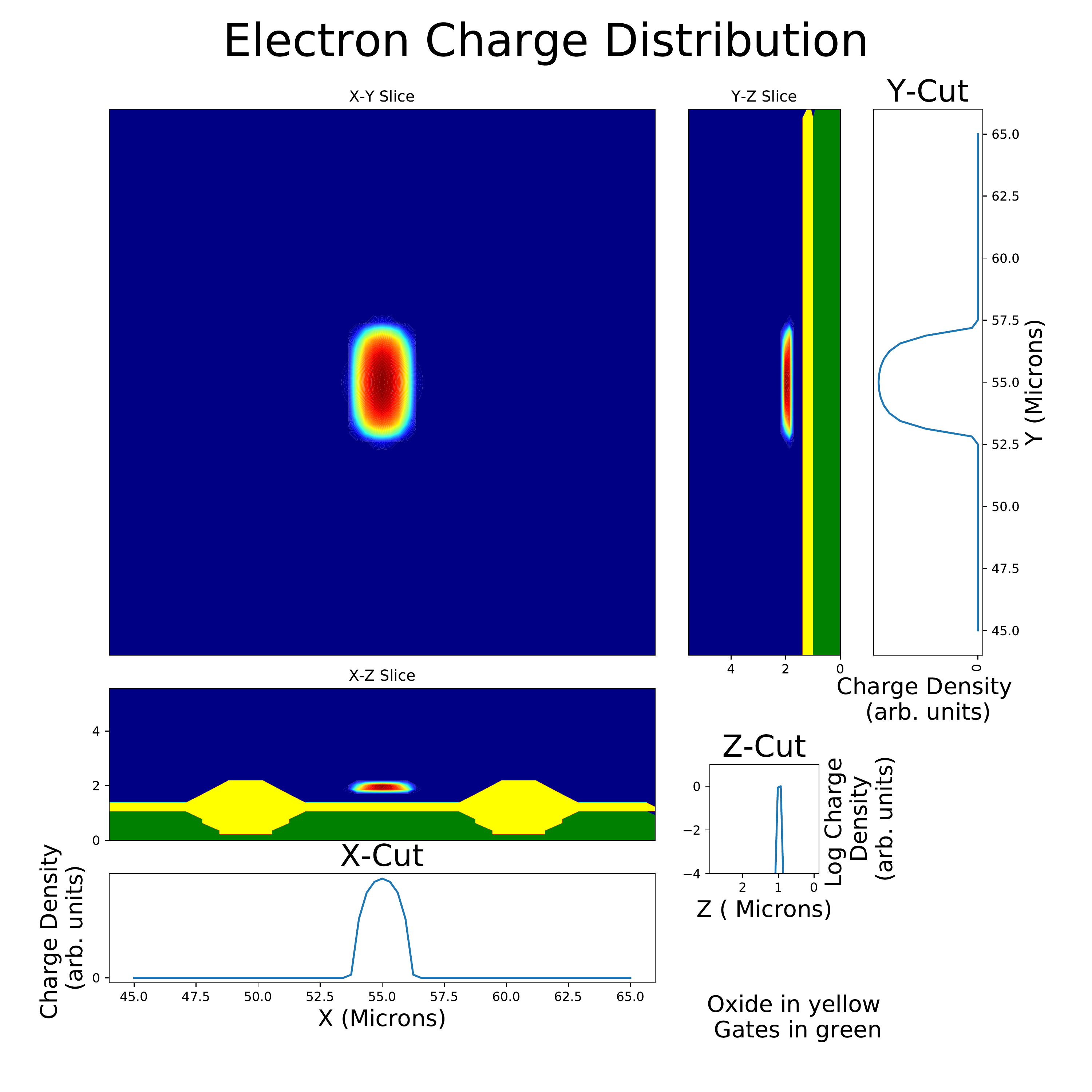}}
	\subfigure[b][40K electrons]{\includegraphics[trim=0.0in 0.0in 0.0in 0.0in,clip,width=0.45\textwidth]{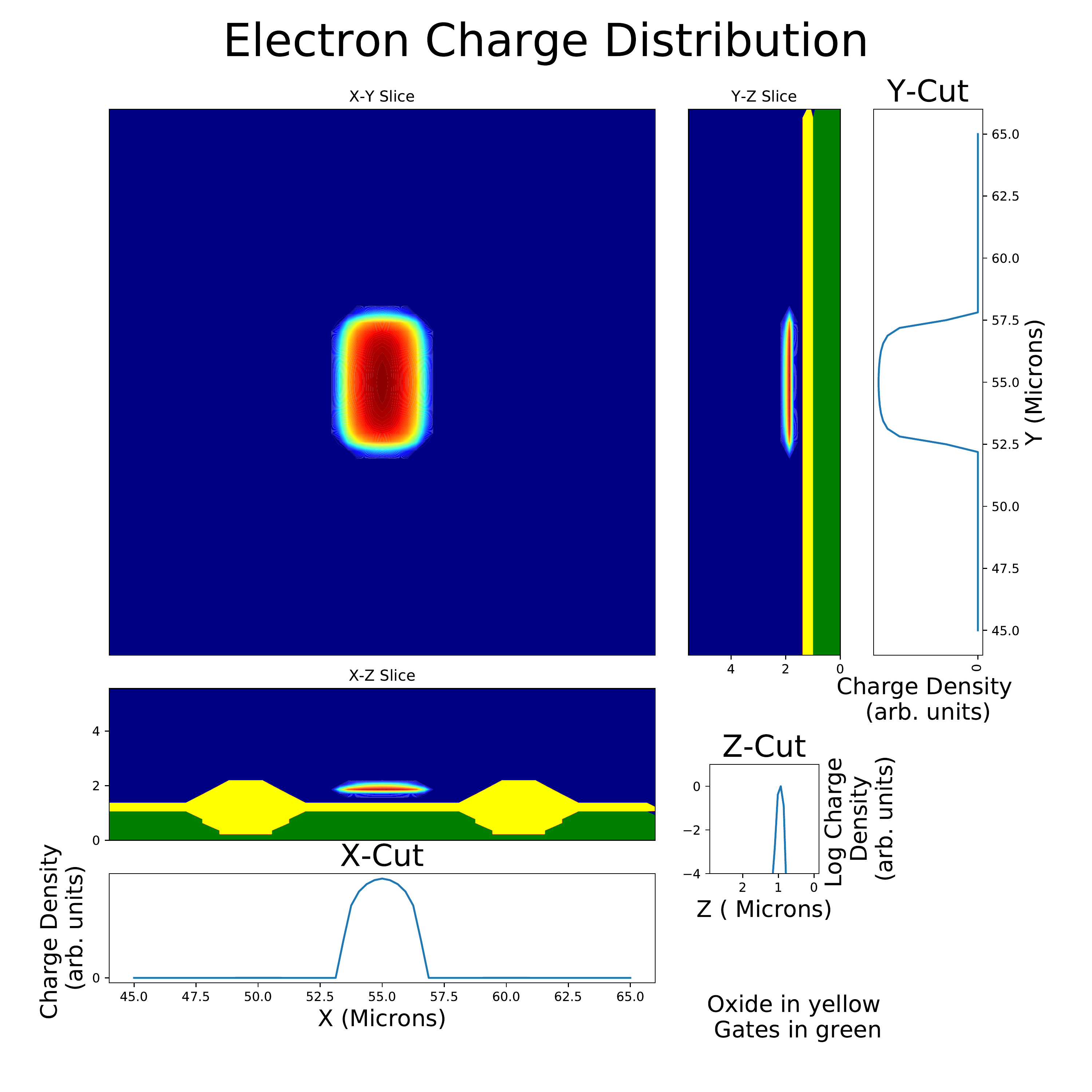}}
	\subfigure[b][70K electrons]{\includegraphics[trim=0.0in 0.0in 0.0in 0.0in,clip,width=0.45\textwidth]{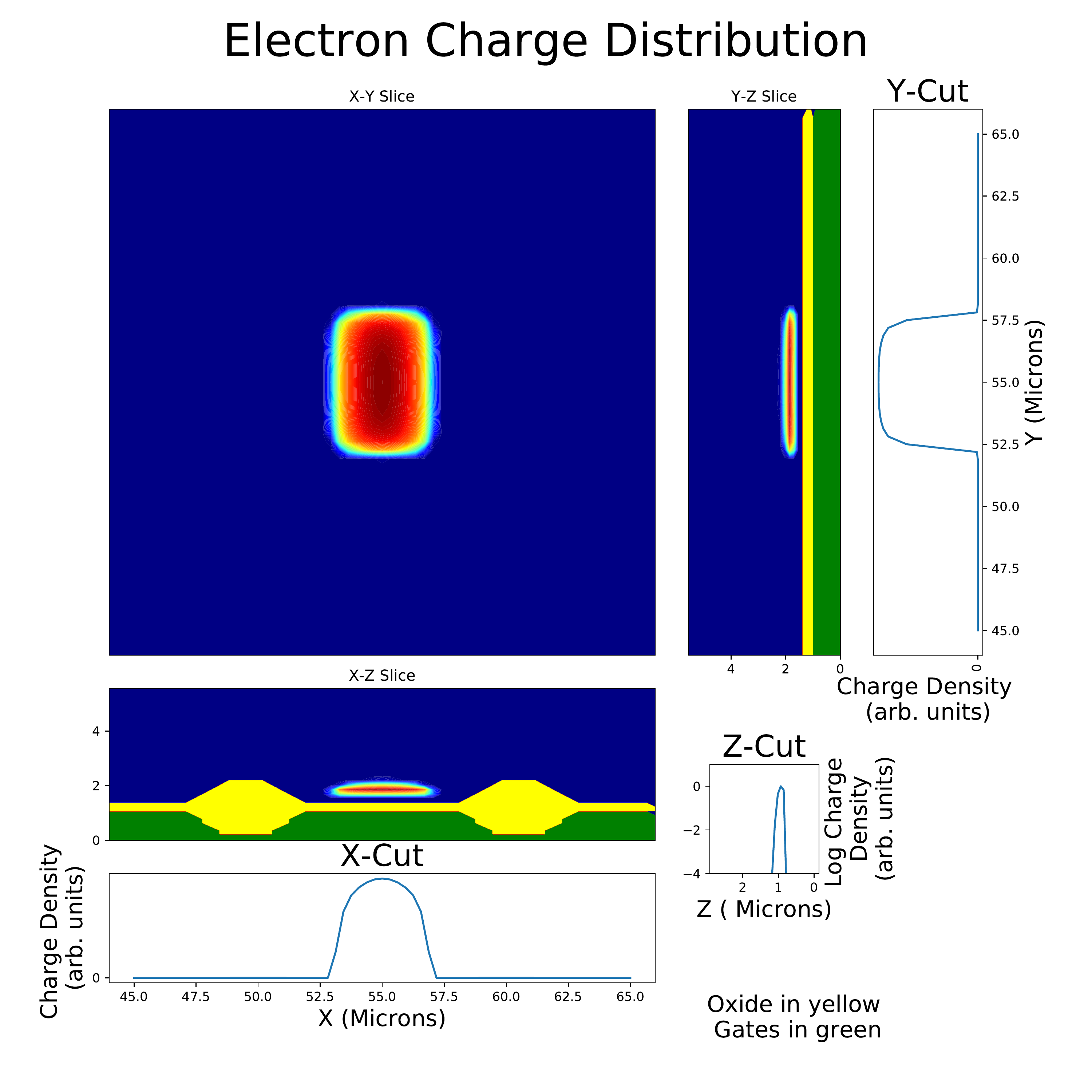}}
	\subfigure[b][100K electrons]{\includegraphics[trim=0.0in 0.0in 0.0in 0.0in,clip,width=0.45\textwidth]{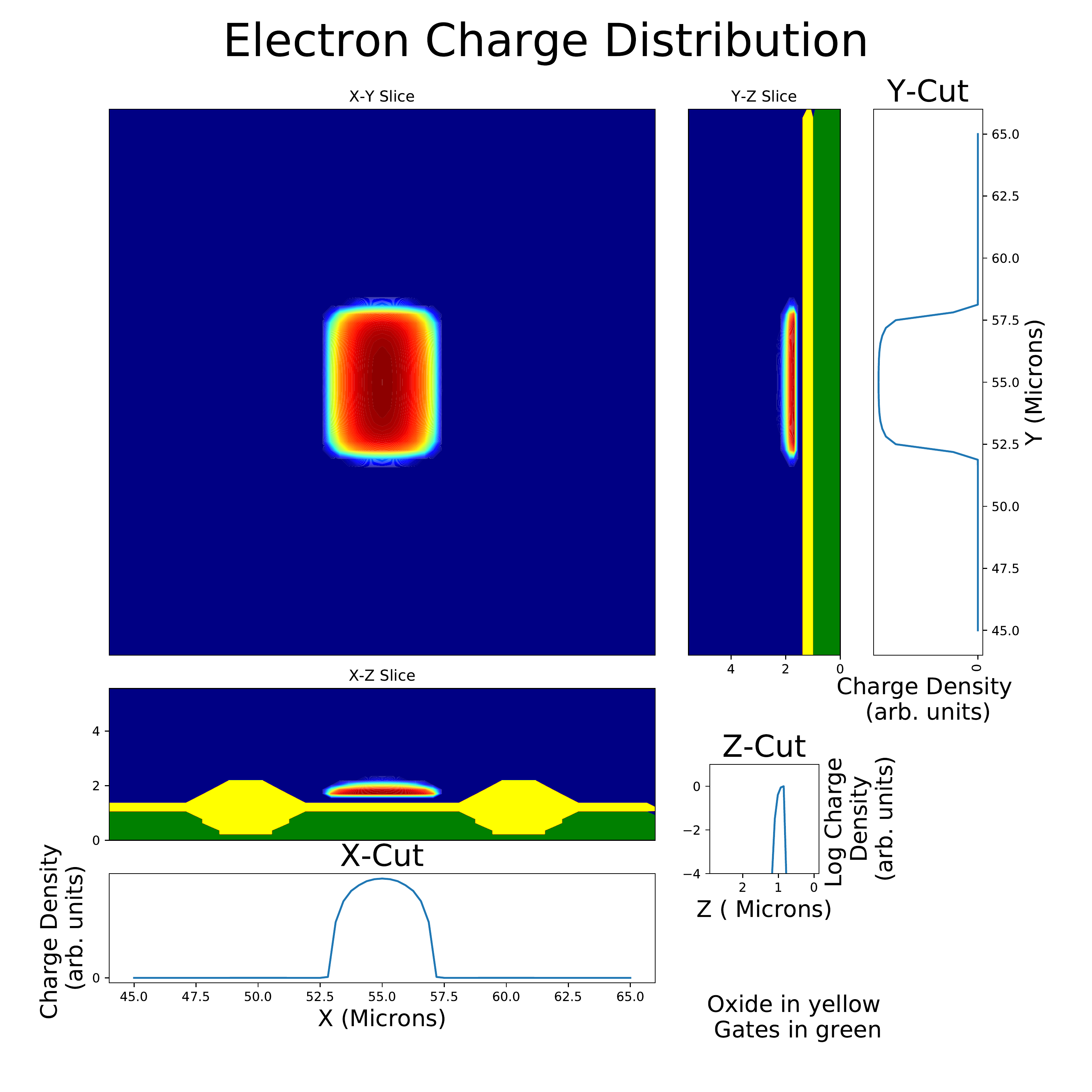}}

  \caption{Simulation of the charge build-up in a single pixel in an ITL CCD.  The Poisson\_CCD code solves self-consistently for the potential in the silicon and the location of the mobile charges (electrons in this case), given the fixed lattice doping and the applied potentials.  We can see here that as more charge is added, the charge cloud spreads out in all three dimensions.  We can then study the impact of this charge re-distribution on the linearity of the BF effect}
  \label{Charge_Build_Up}
  \end{figure}

\begin {figure}
	\centering
	\subfigure[b][Area Distortion vs Flux]{\includegraphics[trim=0.0in 0.0in 0.0in 0.0in,clip,width=0.75\textwidth]{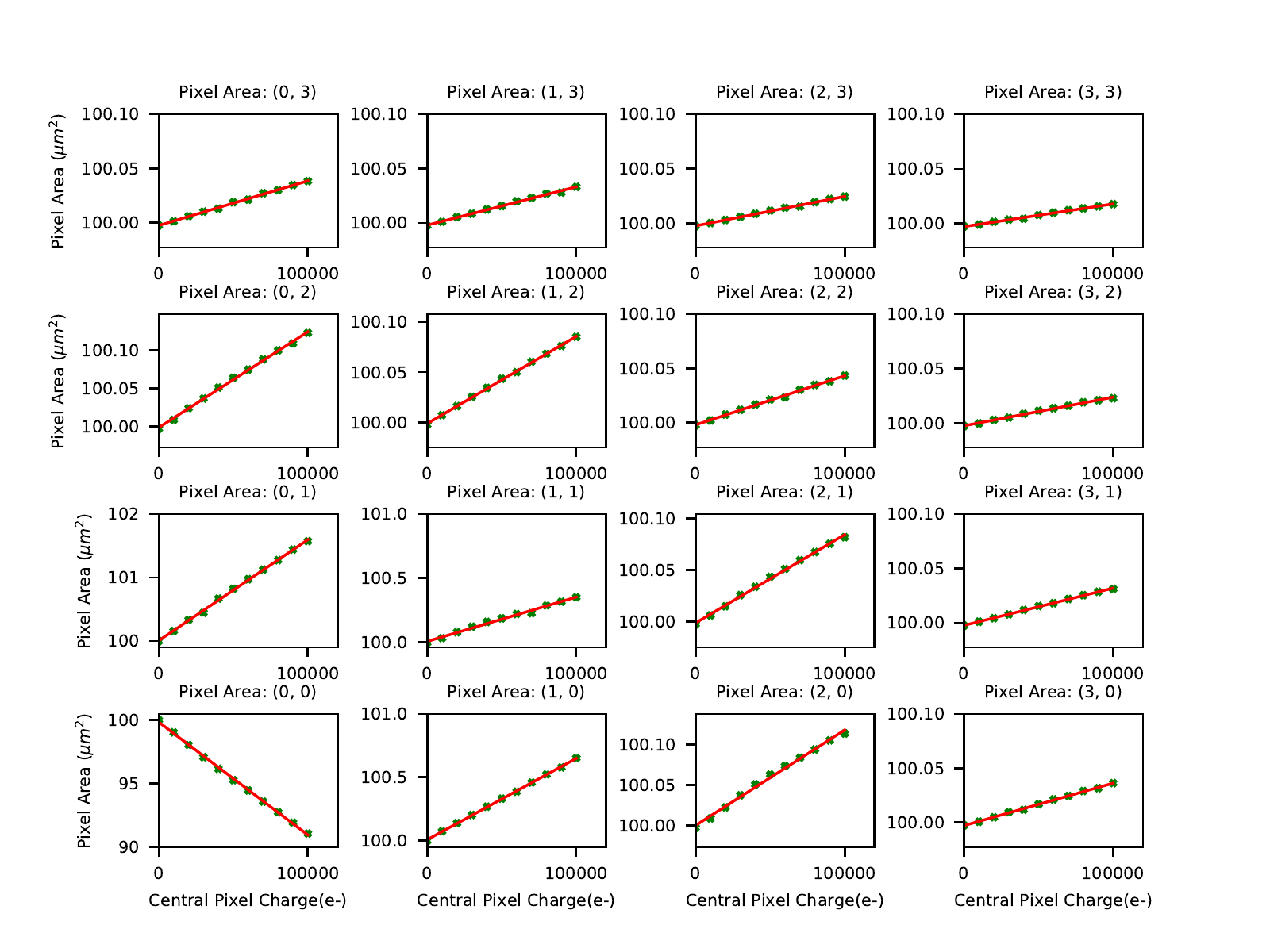}}
	\subfigure[b][Residuals of Linear Fit]{\includegraphics[trim = 0.0in 0.0in 0.0in 0.0in, clip, width=0.75\textwidth]{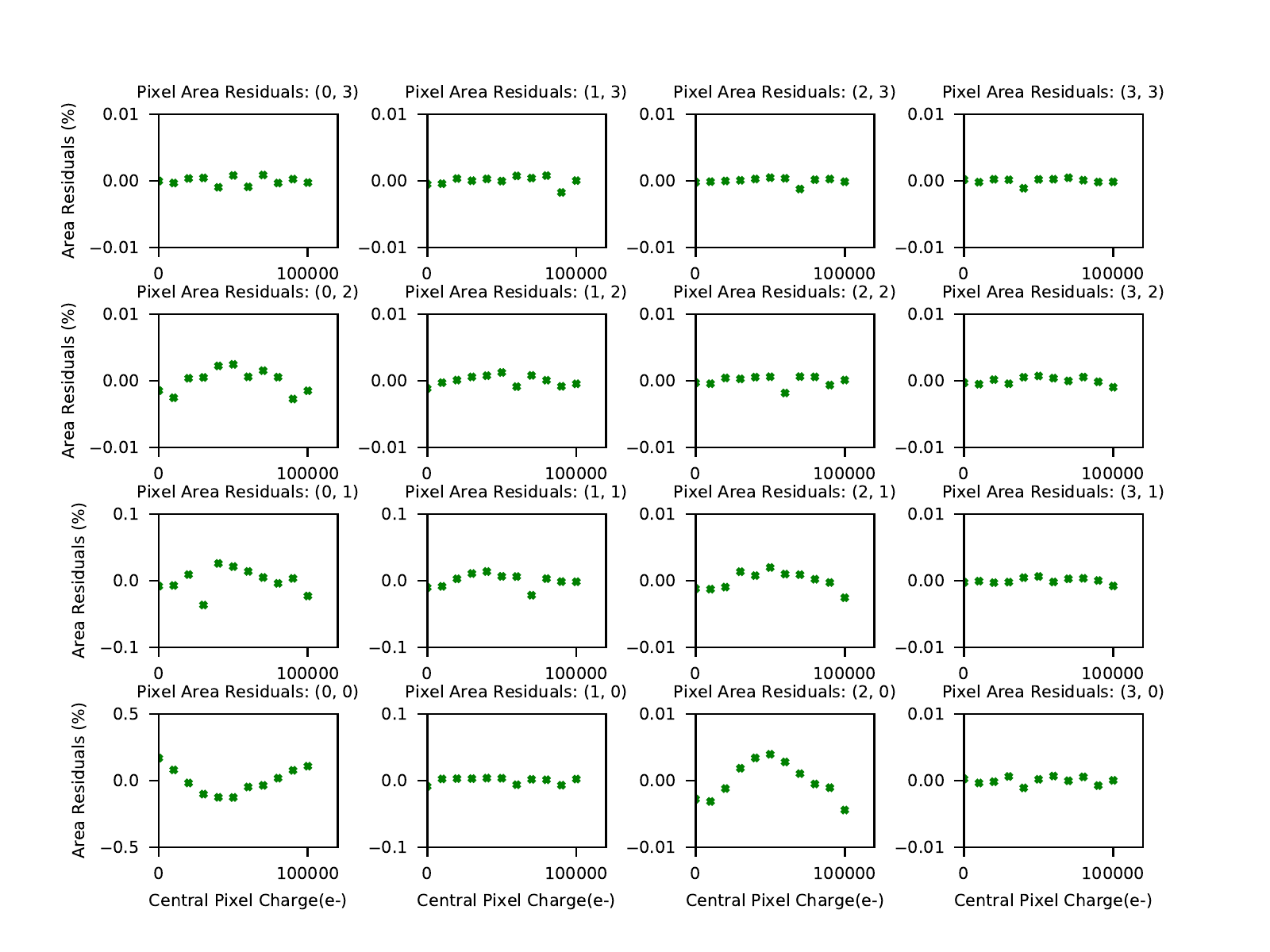}}
  \caption{Simulations of the pixel area distortion as a function of charge in the central pixel.  The central pixel (lower left) loses area, and all of the surrounding pixels gain area.  As can be seen, the trend is quite linear.  The residuals show no trend which would indicate the need for higher order terms.  The small scatter in the simulated areas is due to the finite numerical precision of the binary search algorithm, which has been quantified at about 0.003 $\rm \mu m^2$.}
  \label{Pixel_Area_Linearity}
  \end{figure}

Since the code self-consistently calculates the potential and charge distribution within the CCD volume, we can use the simulations to evaluate the impact of the expanding charge cloud on the pixel distortions.  We expect that the charge cloud will expand in all three dimensions as more electrons are added to the pixel.  This should have some impact on the resulting pixel distortions and hence on the BF effect.  The simulator allows us to quantify this effect.  Figure \ref{Charge_Build_Up} shows the resulting charge cloud, and we can see it expanding as the amount of charge is increased.  Figure \ref{Pixel_Area_Linearity} shows the impact on pixel distortions out to three pixels away from the central pixel.  The trend is quite linear, and the residuals show no systematic effects requiring higher order terms.  The simulations are continued up to the point where the central pixel contains a charge which is approximately 80\% of full well.  We do not expect to use images closer to saturation than this in the modeling of the PSF.

Rasmussen et.al. \cite{rasmussen2016} found significant departures from linearity as the stored charge in the pixel increases, so one is driven to ask what is different between that study and this work.  The simulations done in the Rasmussen study approximate the stored charge distribution as a dipole, effectively concentrating the stored charge at a point in the center of the pixel, whereas this work solves self-consistently for the charge distribution and the potential, thereby giving more realistic charge distributions.  We believe this is what accounts for the difference between the two studies, although this is by no means certain and further study would be illuminating.

\subsection{Linearity of charge superposition}
\begin {figure}[p]
	\centering
	\subfigure[b][Charge distribution]{\includegraphics[trim=0.0in 0.0in 0.0in 0.0in,clip,width=0.40\textwidth]{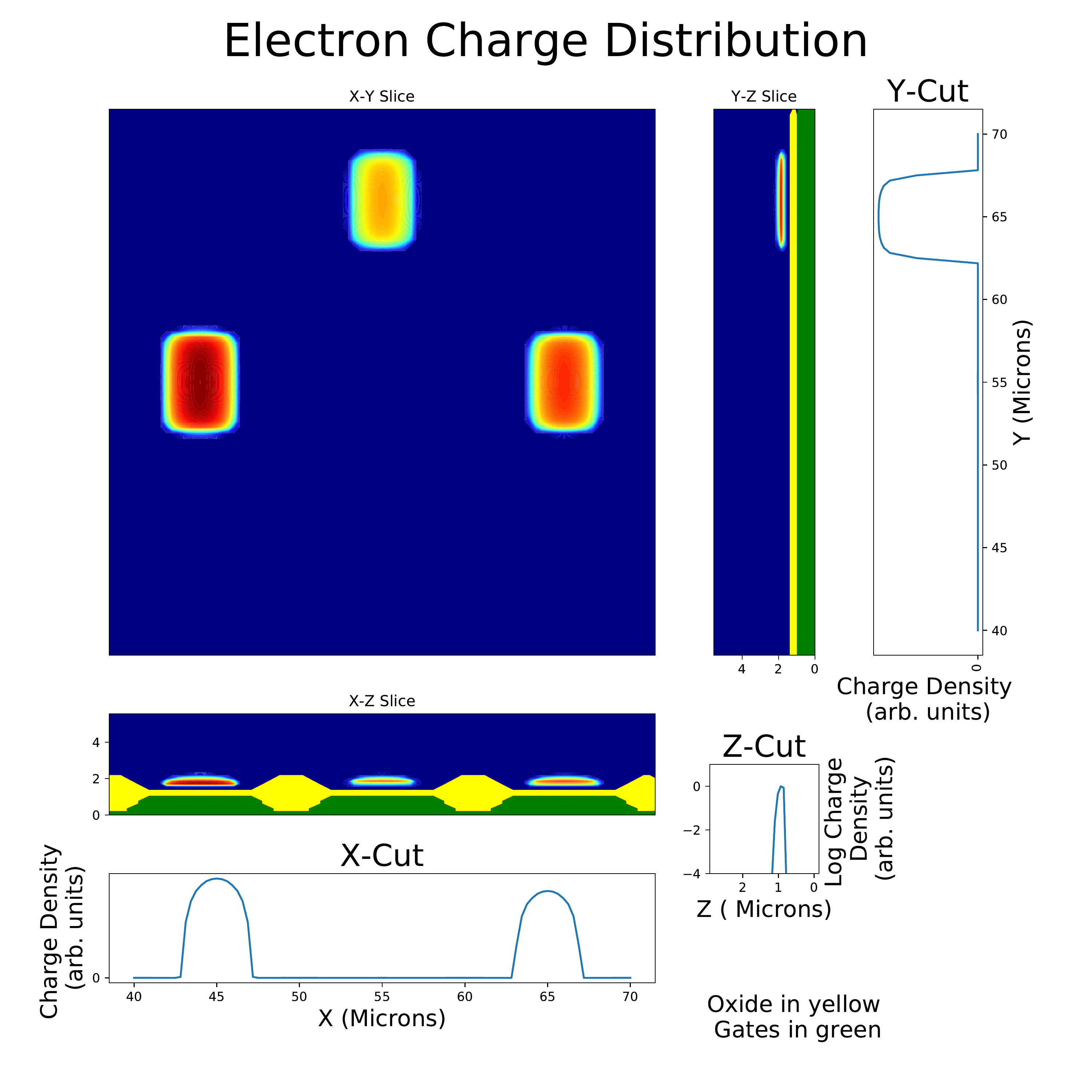}}
	\subfigure[b][Full simulation of pixel distortion]{\includegraphics[trim = 0.0in 0.0in 0.0in 0.0in, clip, width=0.90\textwidth]{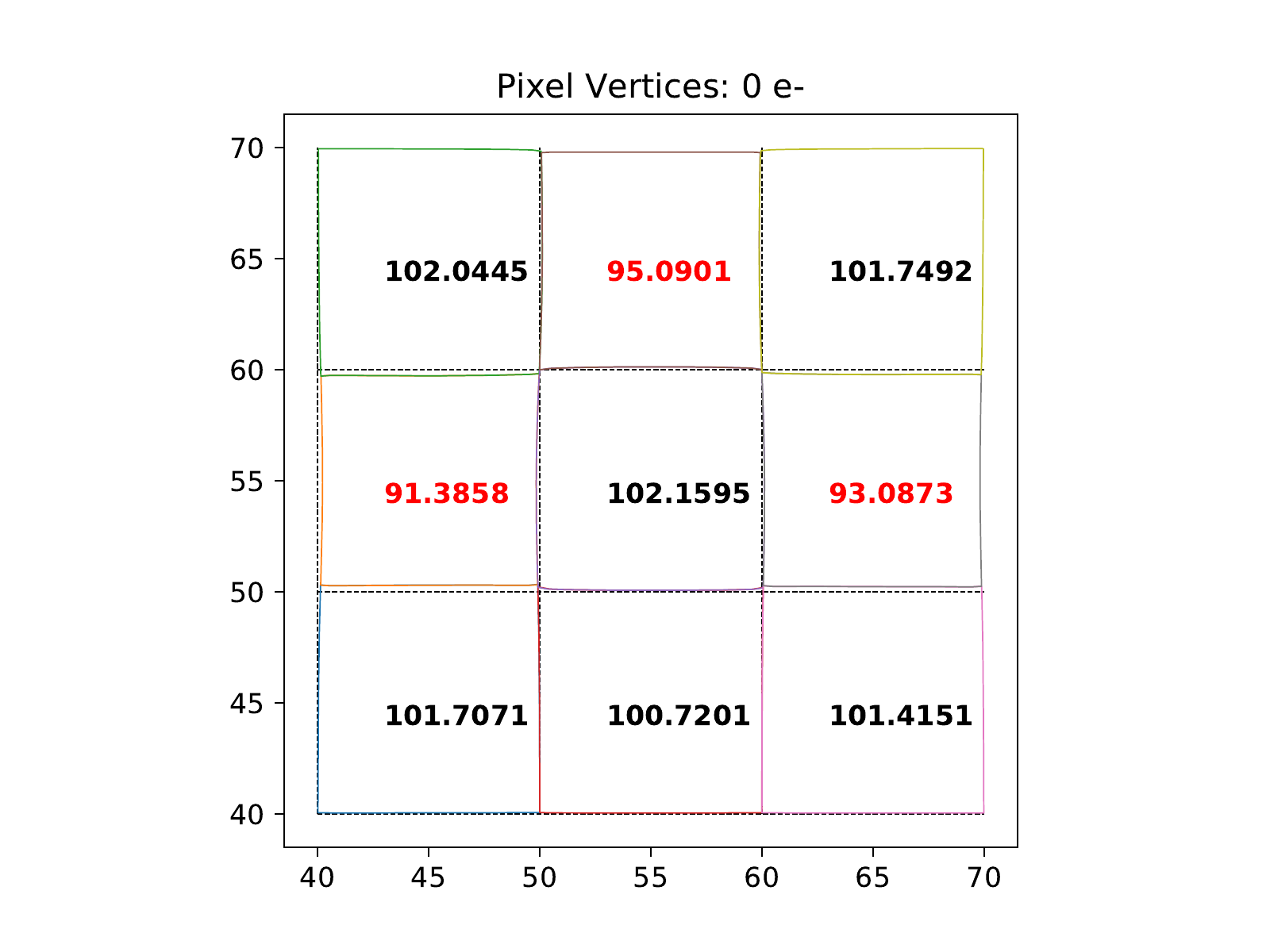}}
  \caption{Test of superposition of pixel distortions.  The top panel shows a simulation with charges in three separate pixels.  The center panel shows a full simulation of the pixel distortions which result from this charge distribution. Figure \ref{Pixel_Superposition_2} shows a comparison of the full simulation to a model of the pixel distortions assuming linear superposition of the three separate charges.}
  \label{Pixel_Superposition_1}
  \end{figure}

\begin {figure}[p]
	\centering
	\subfigure[b][Model of pixel distortions assuming linear superposition.]{\includegraphics[trim = 0.0in 1.0in 0.0in 2.0in, clip, width=1.00\textwidth]{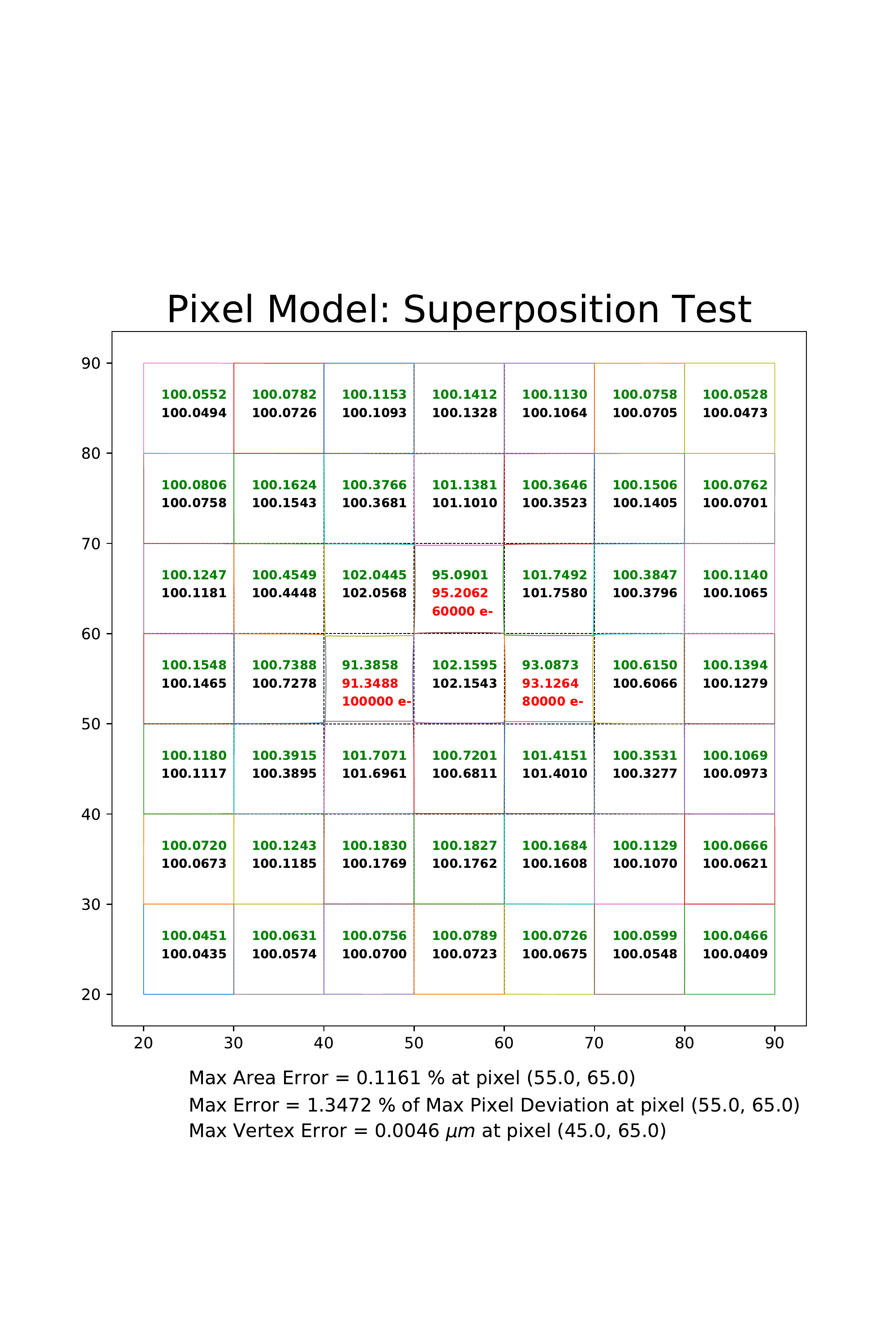}}
  \caption{Test of superposition of pixel distortions.  This shows a comparison of the full simulation to a model of the pixel distortions assuming linear superposition of the three separate charges.  The green areas are the full simulation, and the red and black areas are what result from the superposition assumption.  The worst case error is about 0.1\% of the pixel area, or about 1 \% of the worst case pixel distortion. The worst case pixel vertex error is less than 0.01 $\mu m$ }
  \label{Pixel_Superposition_2}
  \end{figure}

\begin {figure}[p]
	\centering
	\subfigure[b][Model of pixel distortions assuming linear superposition.]{\includegraphics[trim = 0.0in 1.0in 0.0in 2.0in, clip, width=1.00\textwidth]{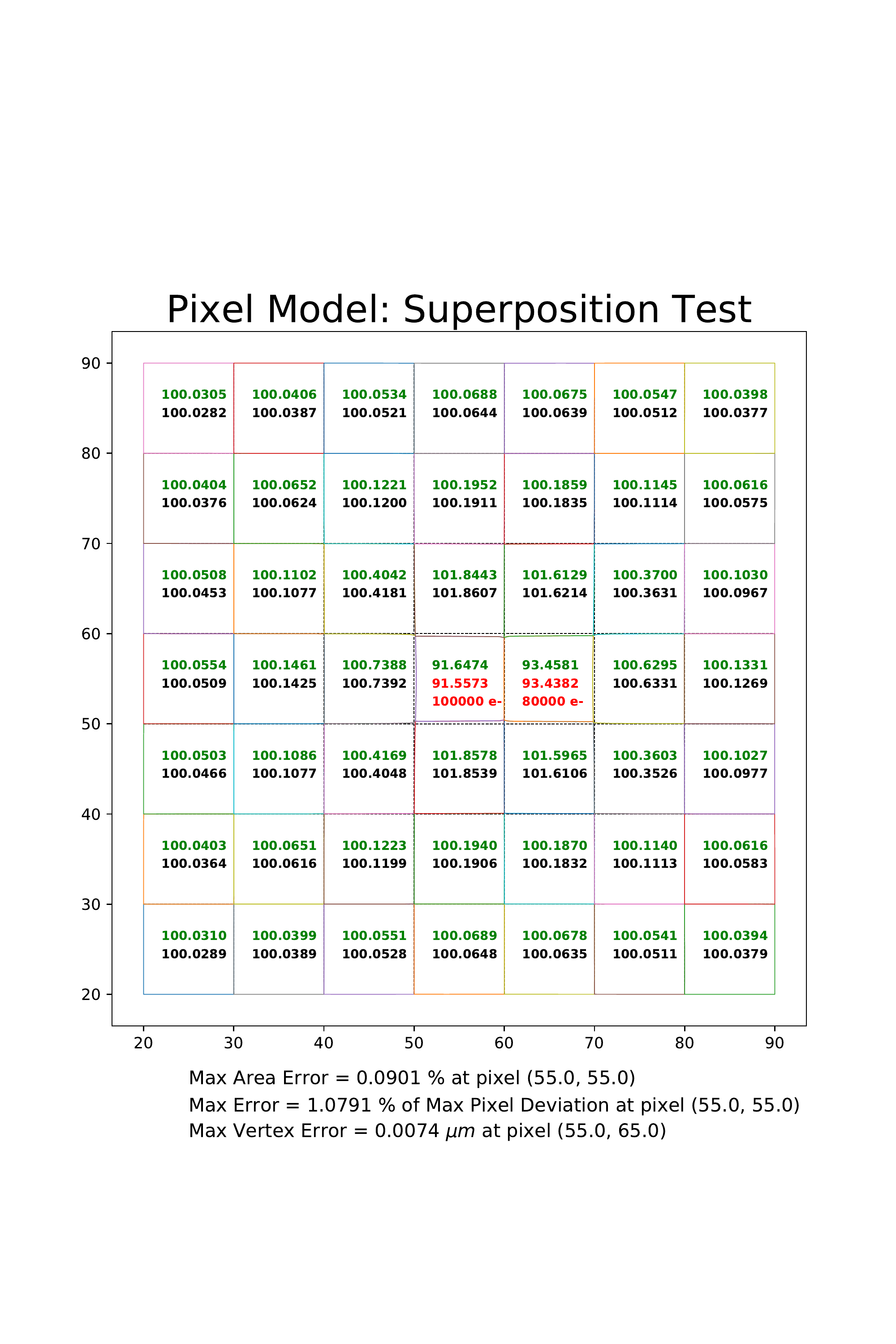}}
  \caption{Test of superposition of pixel distortions.  This shows a comparison of the full simulation to a model of the pixel distortions assuming linear superposition of the three separate charges.  The green areas are the full simulation, and the red and black areas are what result from the superposition assumption.  The worst case error is about 0.1\% of the pixel area, or about 1 \% of the worst case pixel distortion. The worst case pixel vertex error is less than 0.01 $\mu m$ }
  \label{Pixel_Superposition_3}
  \end{figure}

\begin {figure}[p]
	\centering
	\subfigure[b][Model of pixel distortions assuming linear superposition.]{\includegraphics[trim = 0.0in 1.0in 0.0in 2.0in, clip, width=1.00\textwidth]{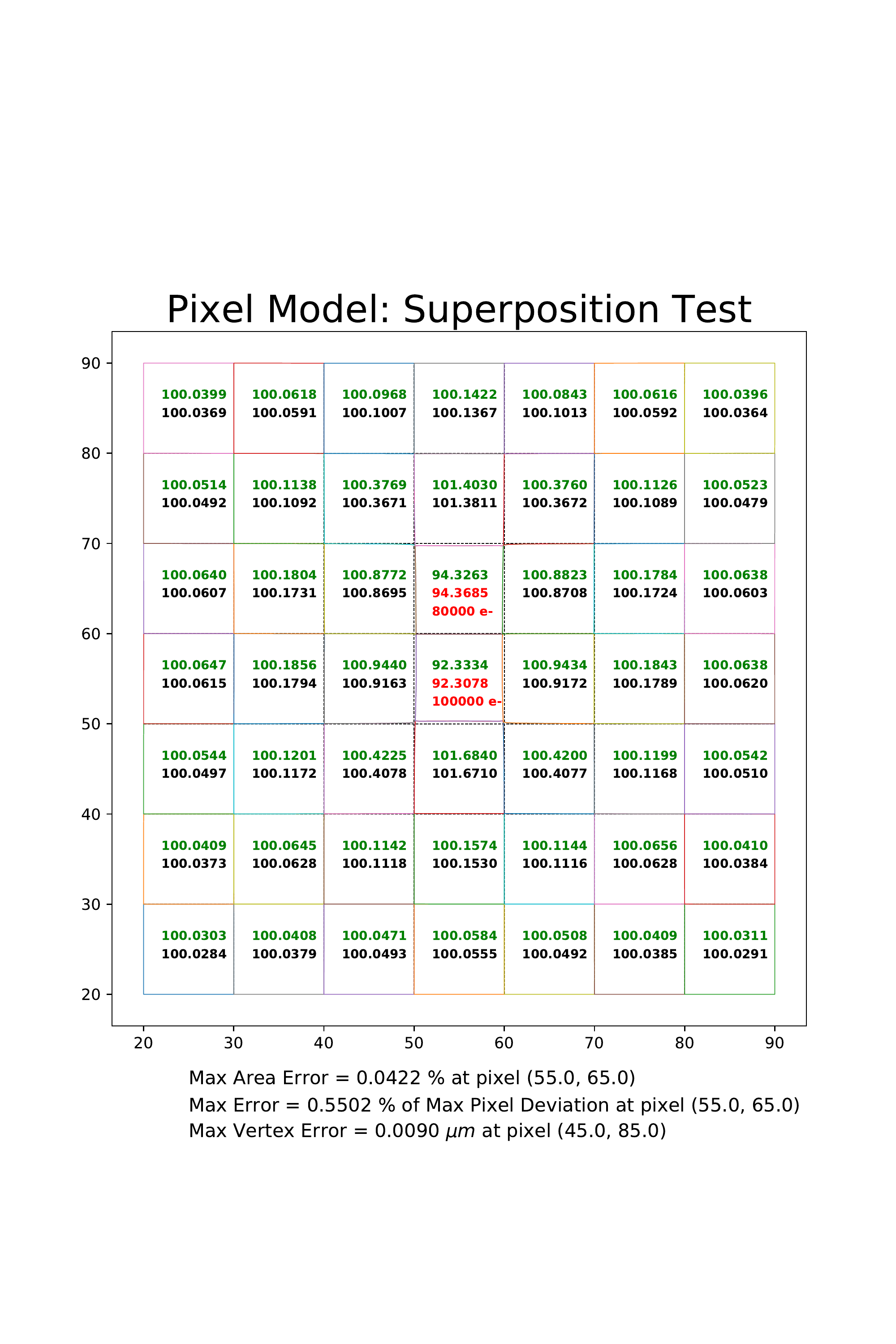}}
  \caption{Test of superposition of pixel distortions.  This shows a comparison of the full simulation to a model of the pixel distortions assuming linear superposition of the three separate charges.  The green areas are the full simulation, and the red and black areas are what result from the superposition assumption.  The worst case error is about 0.1\% of the pixel area, or about 1 \% of the worst case pixel distortion. The worst case pixel vertex error is less than 0.01 $\mu m$ }
  \label{Pixel_Superposition_4}
  \end{figure}

Another question which can be tested with simulations is the linearity of charge superposition.  A key assumption in the BF correction method being used is that distortions due to charge build-up in one pixel and distortions due to charge build-up in a nearby pixel can be linearly superposed.  To test this assumption, we can run a simulation like the one shown in Figure \ref{Pixel_Superposition_1}.  Here charge is added to three pixels and the simulator is used to calculate the resulting pixel distortions.  Since the simulator self-consistently calculates the potential and charge distribution, any impact of one charge distribution on a nearby charge distribution will be modeled.  To test the accuracy of the superposition assumption, we can then compare the simulated pixel distortions to a model of the pixel distortions obtained by adding up the pixel distortion from a single pixel containing charge (as in Figure \ref{Pixel_Distortions}), with the distortions scaled by the amount of charge in each pixel and displaced appropriately.  The result of this comparison for three different charge distributions is shown in Figures \ref{Pixel_Superposition_2}, \ref{Pixel_Superposition_3}, \ref{Pixel_Superposition_4}.  The worst case error in the calculated pixel area is about 0.1\% of the pixel area, or 1\% of the largest pixel area distortion.  The worst case pixel vertex error is less than 0.01 microns.  So we conclude that superposing the pixel distortions is well justified.

\subsubsection{Comparison of simulated pixel distortions and measured pixel-pixel covariances}

\begin {figure}[p]
	\centering
	\subfigure[b][ITL Detector]{\includegraphics[trim=0.5in 0.0in 1.0in 0.0in,clip,width=0.88\textwidth]{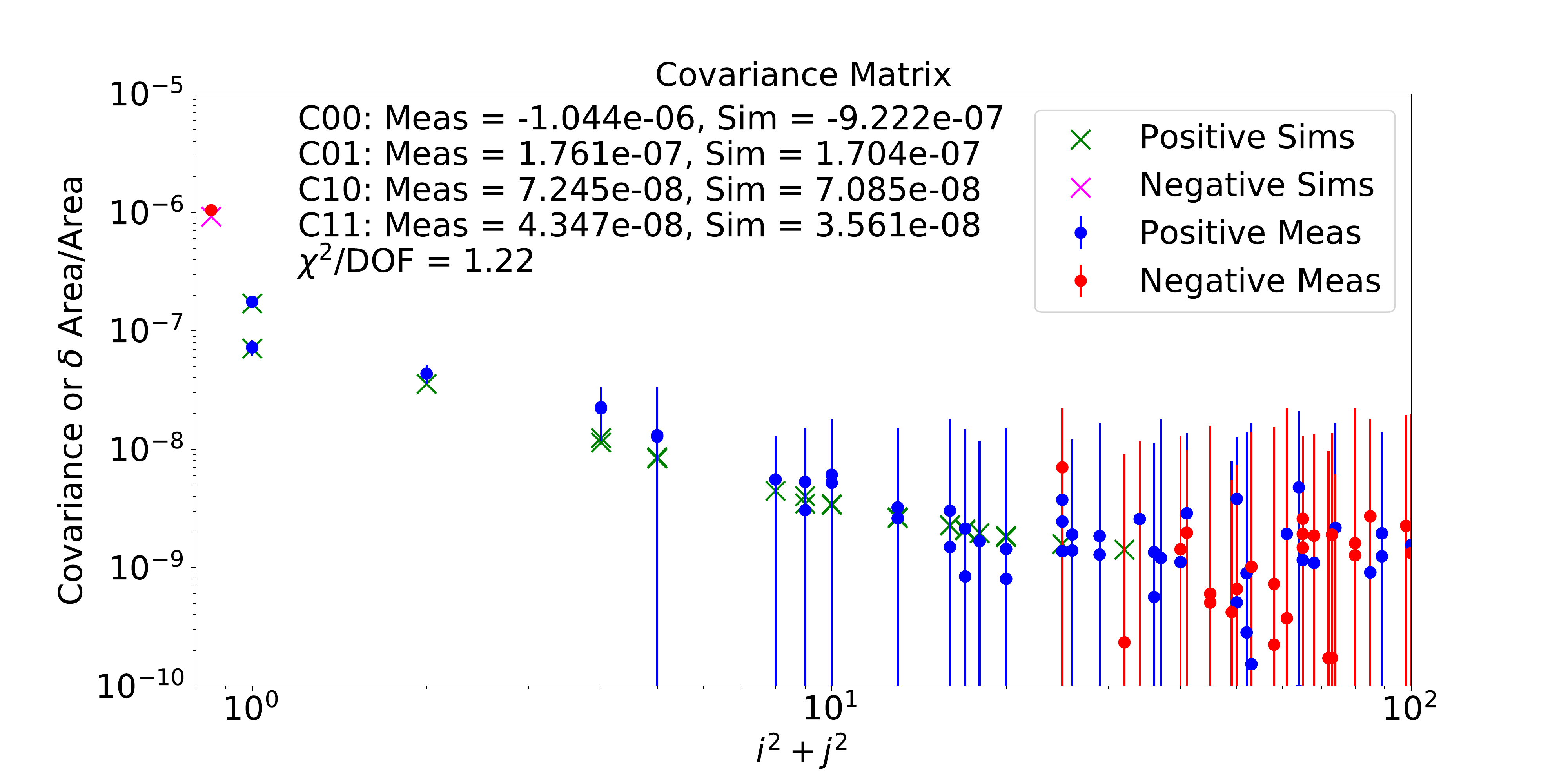}}
	\subfigure[b][E2V Detector]{\includegraphics[trim = 0.5in 0.0in 1.0in 0.0in, clip, width=0.88\textwidth]{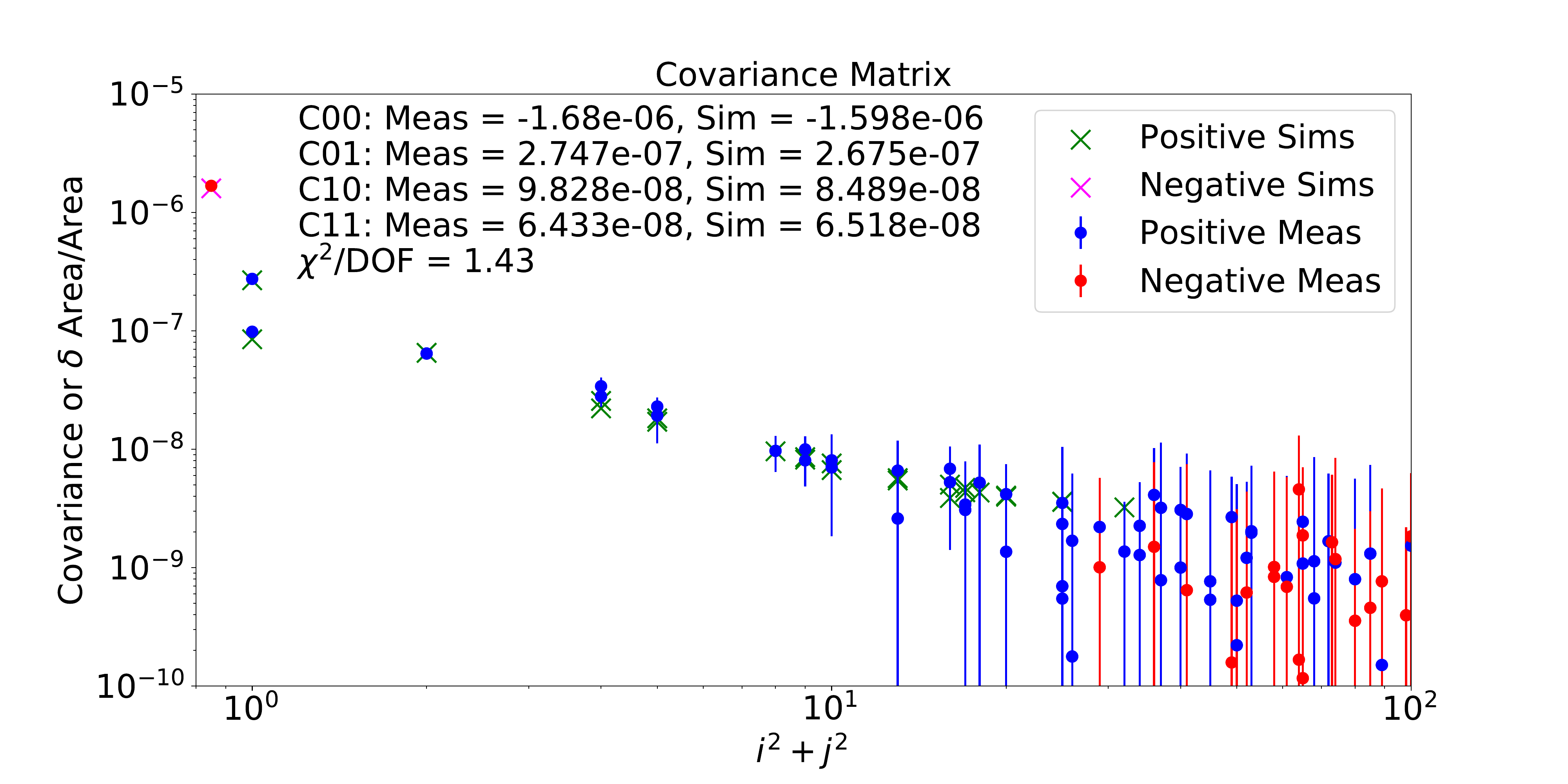}}
  \caption{Covariance measurements and simulations.  The simulated pixel area distortions (see Figure \ref{Pixel_Distortions}) accurately determine the measured pixel-pixel covariances as measured on flat pairs.  The circles are the measured covariances, as extracted by the code in the LSST image reduction pipeline as described in the text.  The crosses are the fractional area distortions as simulated by the Poisson\_CCD code and shown in  Figure \ref{Pixel_Distortions}.  The leftmost point (the central pixel) has been shifted to an X-axis value of 0.8 to allow plotting it on this log-log plot.  Both the E2V and ITL simulations have been informed by physical analysis of both chips, including SIMS dopant profiling and measurements of physical dimensions \cite{CCD-Physical-Analysis}.  Both the covariance measurements and the simulations have been normalized to the distortion caused by one electron.  The asymmetry of the nearest neighbor pixels is correctly modeled, and the simulated values agree with the measurements within the statistical errors.  This lends confidence in using the simulation to study the linearity of the pixel distortions which cause the BF effect.}
  \label{Correlations_Sims}
  \end{figure}

Simulations are fine, but why should we believe these simulations?   As a partial answer to this question, we have compared the simulations of pixel area distortions to measured pixel-pixel covariances extracted from flat pairs.  The flat pairs were measured on the UC Davis LSST beam simulator (\cite{tyson2014}, \cite{Lage_2017}).  The impact of the stored charge on the pixel shapes can be measured by measuring the pixel-pixel covariances on a large number of flat images (\cite{antilogus2014}, \cite{Coulton_2018}).  These covariances are calculated from a large number of flat pairs of varying intensity (see \cite{Coulton_2018} for example) as:

\begin{equation}
C_{i,j} = \frac{\sum_{I,J} (f_{I,J} - \bar{f}) (f_{I+i,J+j} - \bar{f})}{\bar{f}^2(N_{pix} - 1)}
\end{equation}

where $\rm f_{i,j}$ is the difference in flux between the two flats at pixel i,j, and $\rm N_{pix}$ is the number of pixels summed over.  This calculation is implemented in the LSST image reduction pipeline \cite{2018arXiv181203248B}.  Figure \ref{Correlations_Sims} shows the agreement between the measured pixel-pixel covariances on flat field images and the simulated area distortions, as measured and as simulated on LSST CCDs from both CCD vendors.   The agreement is quite good.  The asymmetry of the nearest neighbor pixels is correctly modeled, and the simulated values agree with the measurements within the statistical errors.  This lends confidence in using the simulation to study the second order effects of the pixel distortions which cause the BF effect.

  \section{Non-linearity of covariance measurements}

\begin {figure}[p]
	\centering
	\subfigure[b][Covariances vs Flux]{\includegraphics[trim=0.5in 0.0in 0.5in 0.0in,clip,width=0.99\textwidth]{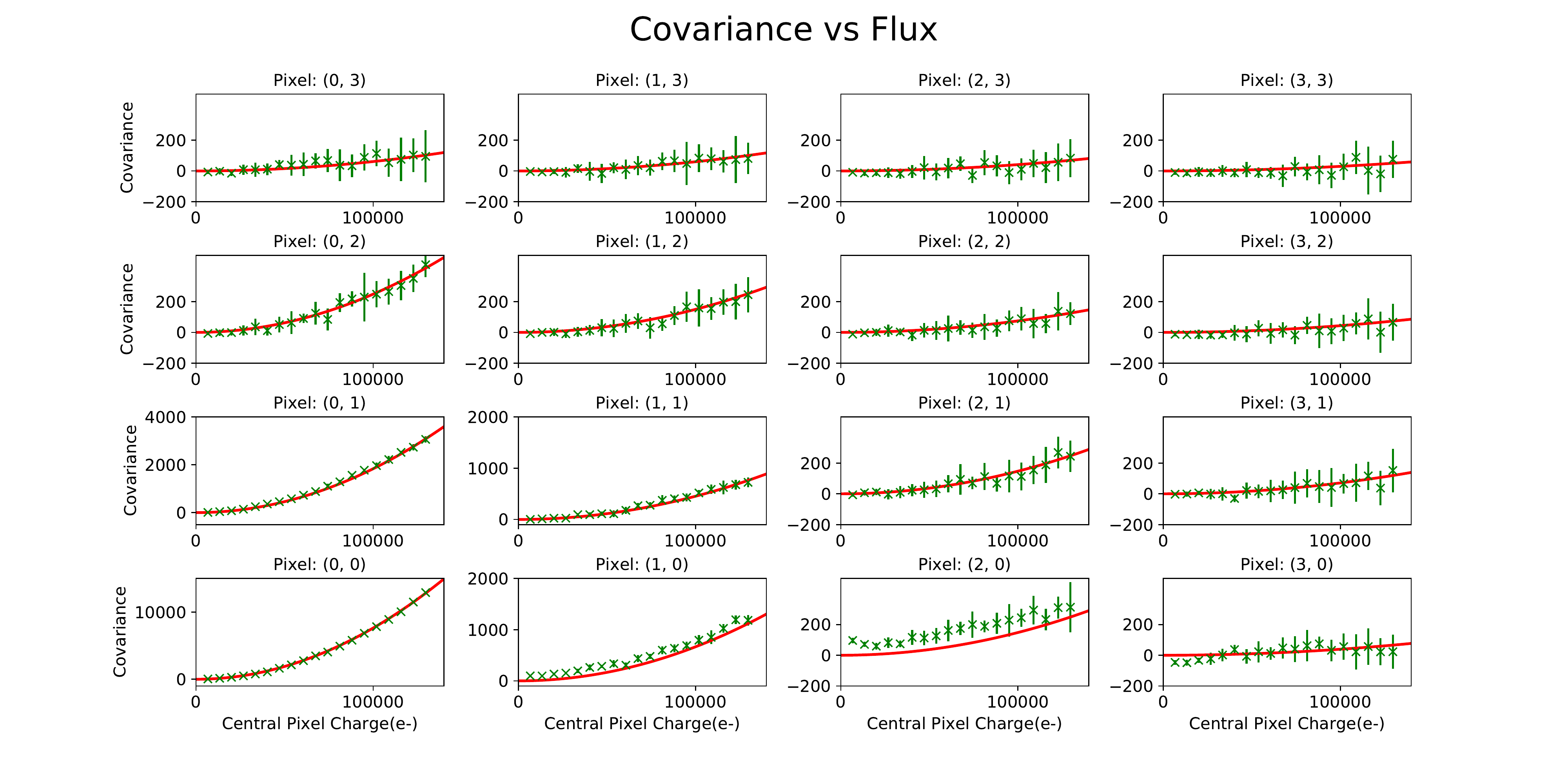}}
	\subfigure[b][Covariances vs Flux]{\includegraphics[trim=0.5in 0.0in 0.5in 0.0in,clip,width=0.99\textwidth]{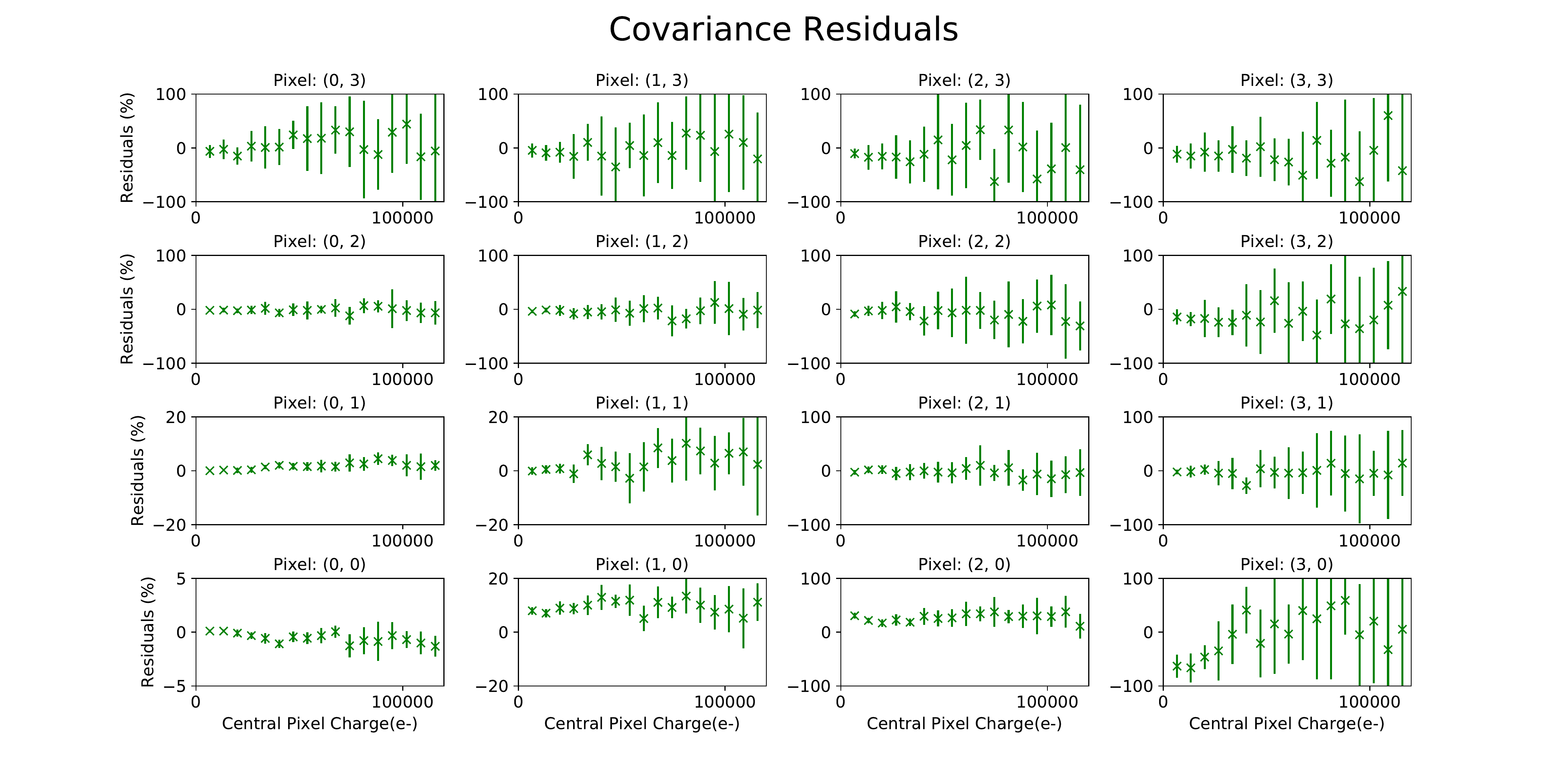}}
  \caption{Quadratic dependence of the covariances for the ITL detector. This is based on 190 flat pairs.  The red line is a quadratic fit - i.e. $\rm C_{ij} = \alpha f^2$, where the value of $\alpha$ is determined to minimize the sum of squared differences.  The residuals are the percentage difference between this quadratic fit and the data (green crosses).    The quadratic fit fits the data well, and there is no apparent systematic trend in the residuals indicating the need for higher order terms.    The largest residuals are in the serial direction, which we believe based on other data are due to serial charge transfer inefficiency (CTI).}
  \label{Correlations_Quad_ITL}
  \end{figure}

  This section looks at departures from linearity of pixel-pixel covariances extracted from flat images made with LSST detectors.  By measuring a series of flats at increasing intensities, we can measure how the pixel-pixel covariances increase as the number of electrons per pixel increases.  As explained in  \cite{antilogus2014} and \cite{Coulton_2018}, these covariances are a direct measure of the electrostatic repulsions that give rise to the BF effect.  Figures \ref{Correlations_Quad_ITL} and \ref{Correlations_Quad_E2V} show the results of these measurements.  Each is based on approximately 100 flat pairs, and the covariances are as described in the preceding section.  To first order, we expect these covariances to increase quadratically with flux.  It can be seen in these figures that the measured covariances are closely fit by a curve proportional to the square of the flux.  The residuals show no systematic trend requiring higher order terms.  The largest residuals are in the serial direction for the ITL device, where it can be seen that most or all of the points are above the quadratic fit line.   We believe, based on other measurements, that these increased covariances arise from a known problem serial charge transfer inefficiency (CTI).  Adequately removing the effect of CTI from the covariance measurements is an area needing more study, and we expect that doing this properly will improve the correction of the BF effect.

\section{BF correction using the Coulton kernel method.}
This section discusses work we have done to check how well the Coulton kernel method corrects measurements of simulated stars measured on the LSST detectors using the UC Davis LSST optical simulator (\cite{tyson2014}, \cite{Lage_2017}).  It should be emphasized that this is still work in progress, and several groups are actively working to evaluate and improve the BF correction.  However, we have made significant progress and we are successfully correcting the majority of the BF effect.  To evaluate the efficacy of the correction, we use the following sequence:
\begin{itemize}
  \item Measure a series of flat pairs with a range of different intensities.  We have typically used 20 different linearly spaced intensities, with the highest intensities being approximately 80\% of full well.
  \item Use the LSST image reduction pipeline to extract the BF kernel from these flats.  For this work, this is done on a single amplifier, which is approximately 500 x 2000 pixels.
  \item Measure a series of spot images of linearly increasing intensities, again with the highest intensities being approximately 80\% of full well.
  \item Characterize the spot sizes as a function of intensity, again using the LSST image reduction pipeline to extract the second moments, both with and without correction.  The plots shown below again use a single amplifier.  Approximately 1000 spots are measured, and the plots below show the mean and 1 sigma of the second moment values.
\end{itemize}

\begin {figure}[p]
	\centering
	\subfigure[b][ITL photon transfer curve - Gain=4.411]{\includegraphics[trim=0.0in 0.0in 0.0in 0.0in,clip,width=0.49\textwidth]{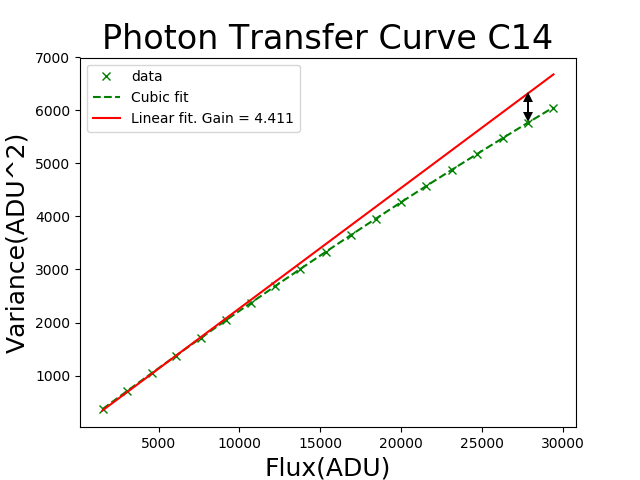}}
	\subfigure[b][E2V photon transfer curve - Gain=4.624]{\includegraphics[trim=0.0in 0.0in 0.0in 0.0in,clip,width=0.49\textwidth]{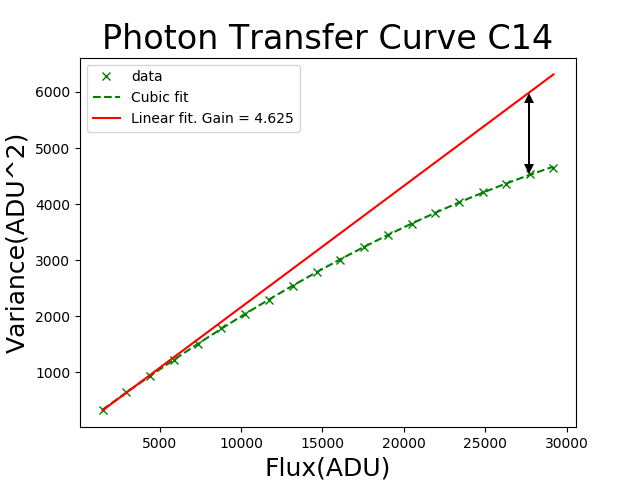}}
  \caption{Photon transfer curves for the two detectors.  The $\rm C_{00}$ covariance term in the covariance matrix which is inverted to obtain the BF correction kernel is the small difference between the linear part of the photon transfer curve and the observed photon transfer curve at high flux, indicated by the black arrows.  Since a small error in the gain determination will lead to a large error in this $\rm C_{00}$ term, it is crucial to get the gain right in order for the correction kernel to be correct.}
  \label{Gains}
  \end{figure}

\begin {figure}[p]
	\centering
	\subfigure[b][E2V Baseline]{\includegraphics[trim=1.0in 0.0in 1.5in 0.0in,clip,width=0.49\textwidth]{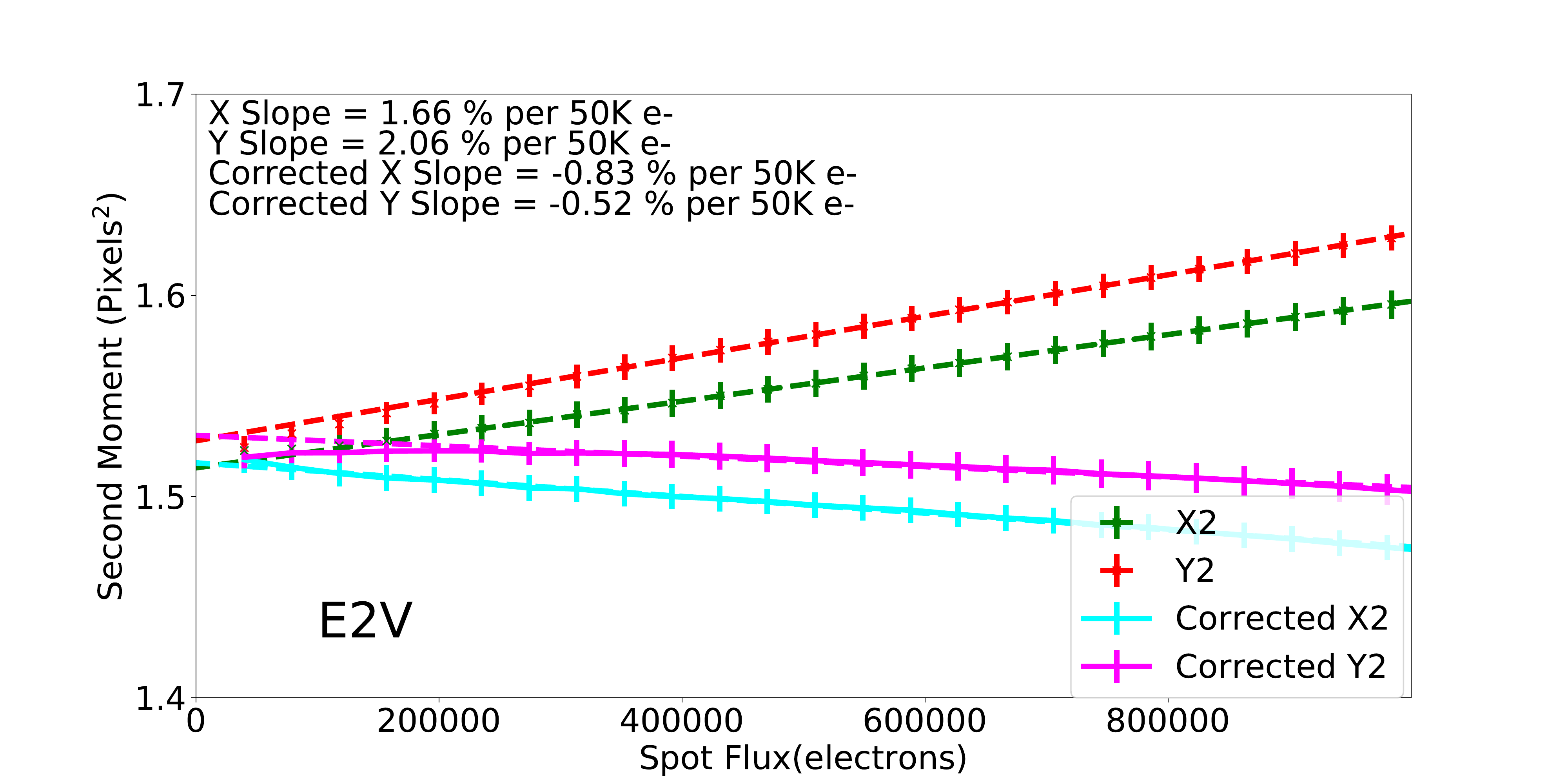}}
	\subfigure[b][E2V Gain]{\includegraphics[trim=1.0in 0.0in 1.5in 0.0in,clip,width=0.49\textwidth]{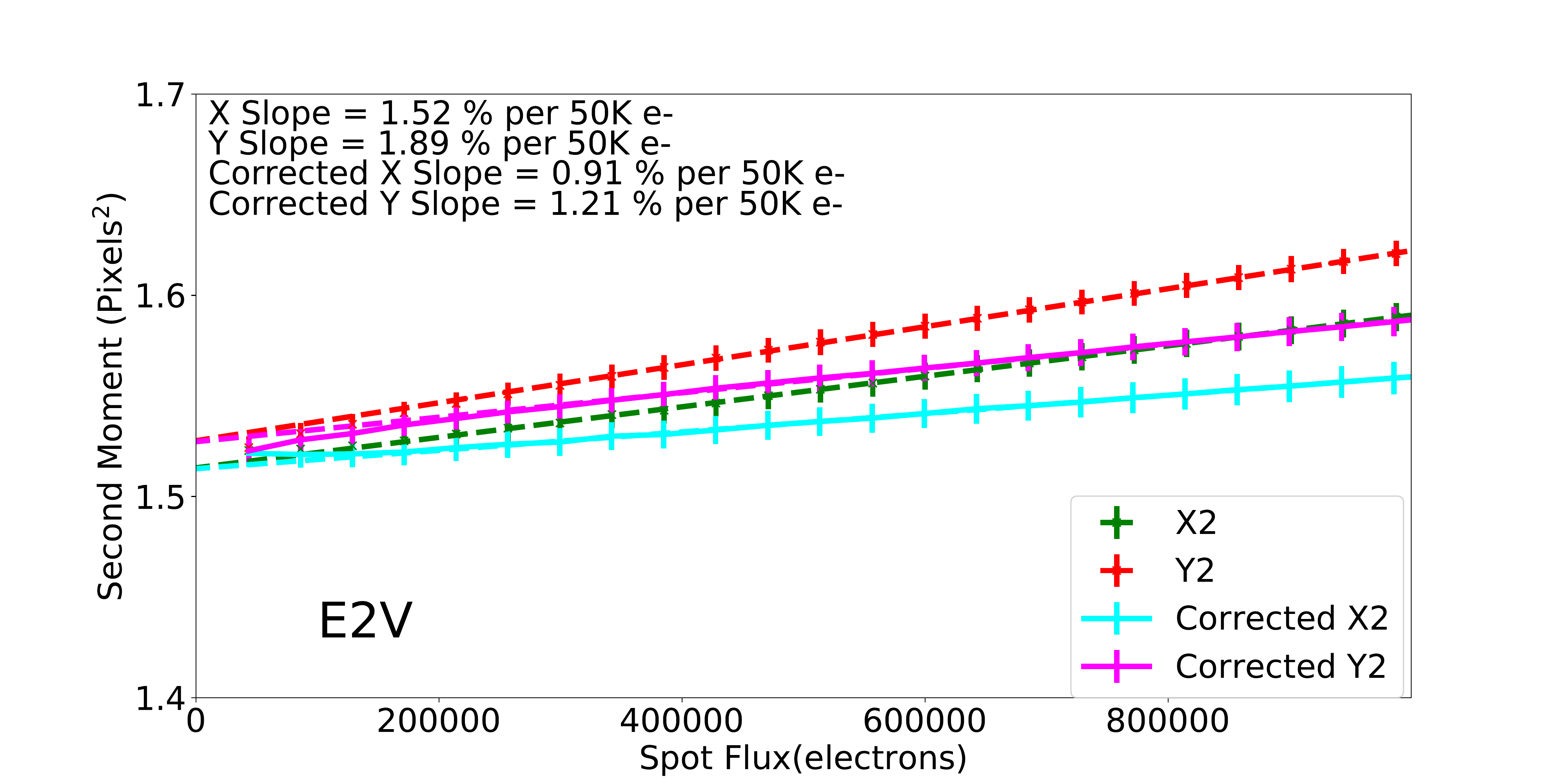}}
	\subfigure[b][E2V Gain Quad]{\includegraphics[trim=1.0in 0.0in 1.5in 0.0in,clip,width=0.49\textwidth]{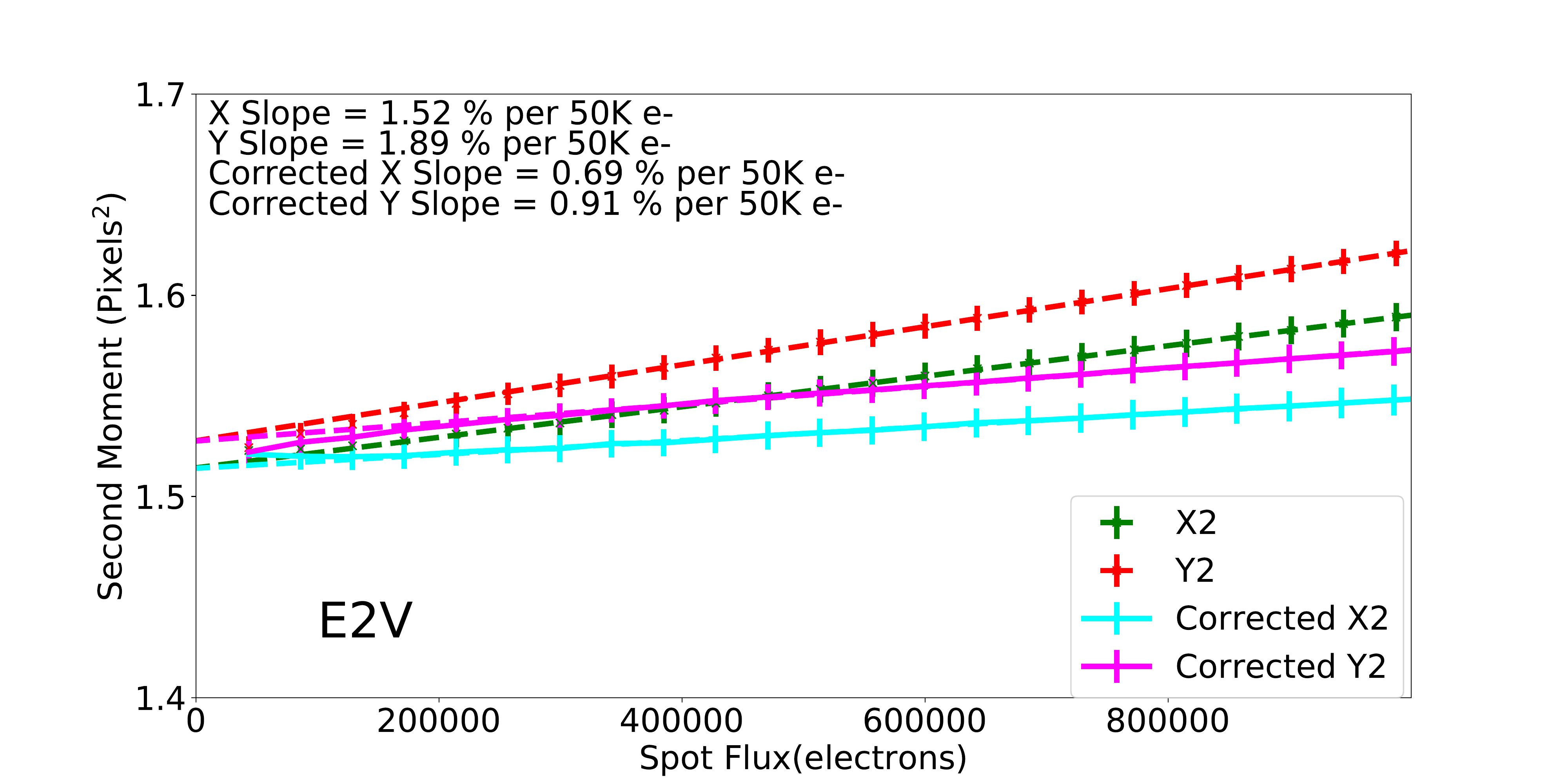}}
	\subfigure[b][E2V Gain Quad Zero]{\includegraphics[trim=1.0in 0.0in 1.5in 0.0in,clip,width=0.49\textwidth]{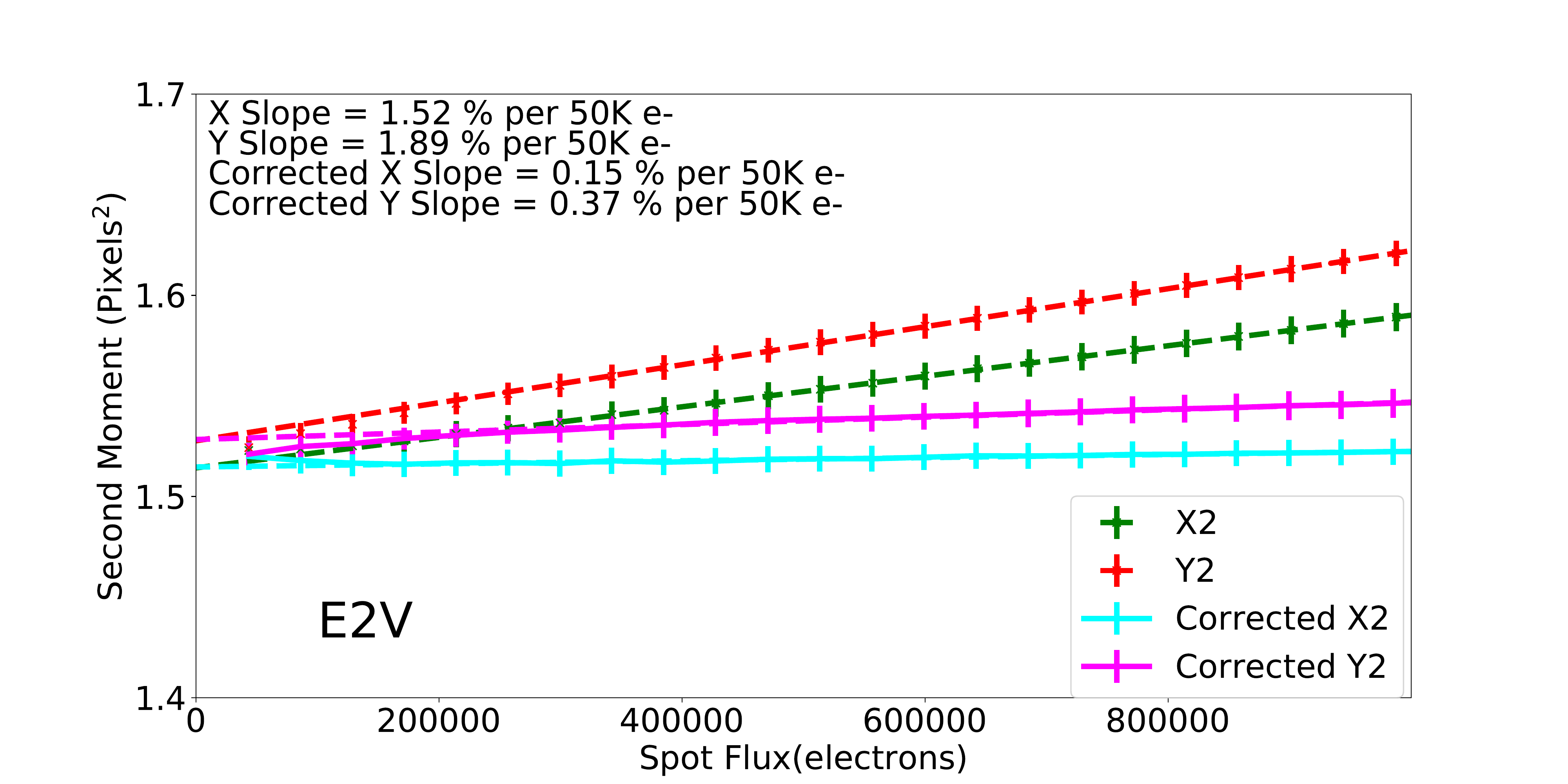}}
	\subfigure[b][E2V Gain Quad Zero Model]{\includegraphics[trim=1.0in 0.0in 1.5in 0.0in,clip,width=0.49\textwidth]{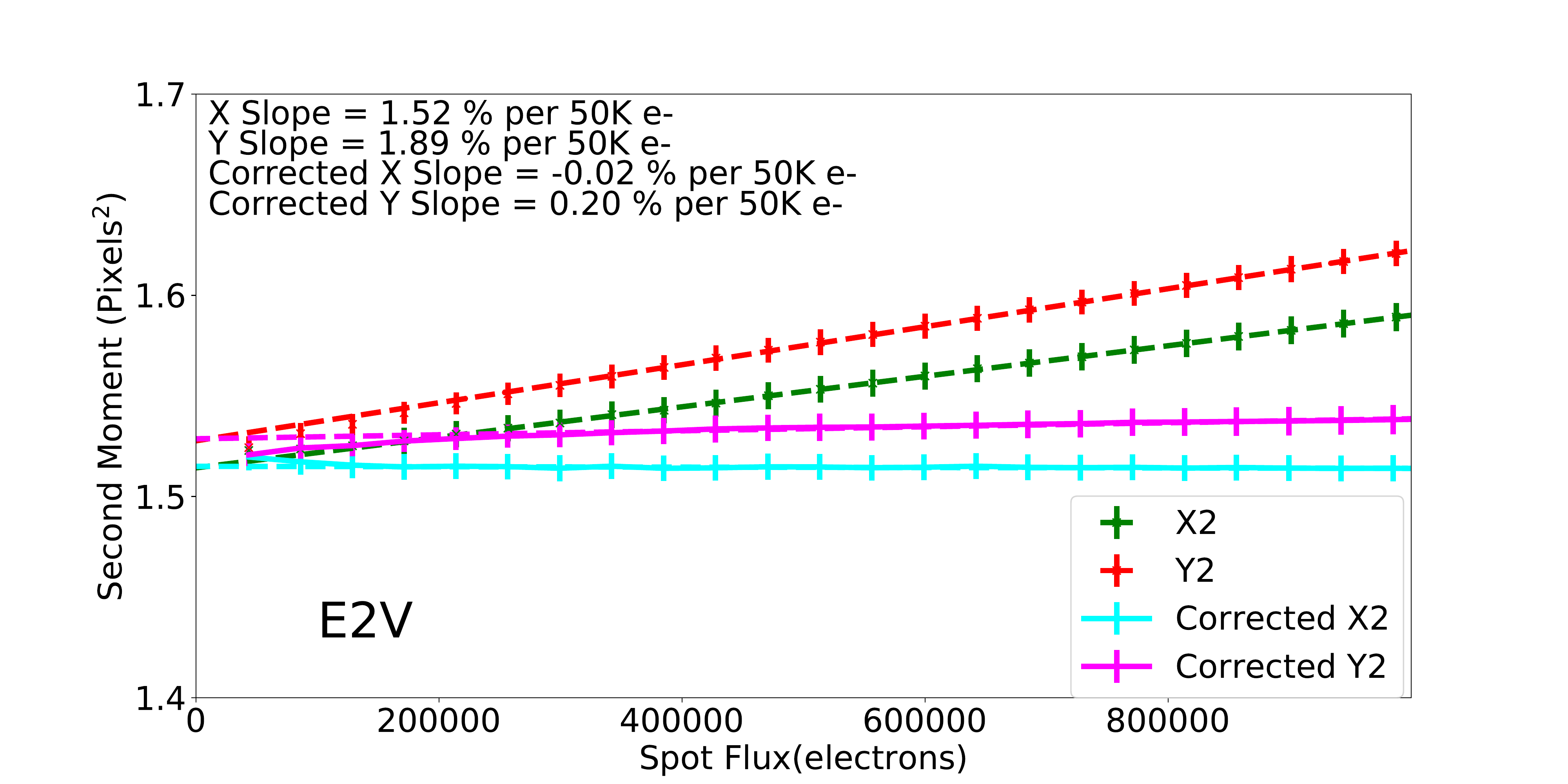}}        

  \caption{Summary of BF correction results on the E2V detector using the code improvements described in this work, with each code improvement added sequentially.  ``Baseline'' is the existing code.  ``Gain'' corrects the gain determination.  ``Quad'' uses the quadratic fit to the covariances instead of simple averaging.  ``Zero'' adjusts the $\rm C_{00}$ covariance term to force the covariance matrix to have zero sum.  ``Model'' uses a fit to the covariance terms for pixels which are 3 pixels or more away from the central pixel, as well as adding a term to account for summing the covariance matrix to infinity.}
  \label{BF_Corrections_E2V}
  \end{figure}

Initial results of this process showed a significant over-correction, which can be seen in the 'E2V Baseline' panel of Figure \ref{BF_Corrections_E2V}.  This was quickly tracked down to an error in the gain calculation.  Figure \ref{Gains} shows Photon Transfer Curves (PTC) for both detectors.  As this figure shows, the $\rm C_{00}$ covariance value is basically the difference between the linear growth of the Poisson variance and the measured variance.  When the covariance matrix is inverted to obtain the correction kernel, as explained in Coulton, et.al., the $\rm C_{00}$ term, because it is the largest term, has a major impact on the BF correction.  For this reason, getting the gain value correct is crucial to successfully correcting the BF effect.  This will be discussed more below.  However, after correcting the gain error, we found the algorithm to now be undercorrecting.  This can be seen in the 'E2V Gain' panel of Figure \ref{BF_Corrections_E2V}.   To improve the BF correction, we have four proposed changes to the kernel extraction method, which result in significant improvement to the correction.  These changes are summarized as follows, where each proposed improvement is given a name for reference in the attached plots.

\begin {figure}[p]
	\centering
	\subfigure[b][Covariances vs Flux]{\includegraphics[trim=0.5in 0.0in 0.5in 0.0in,clip,width=0.99\textwidth]{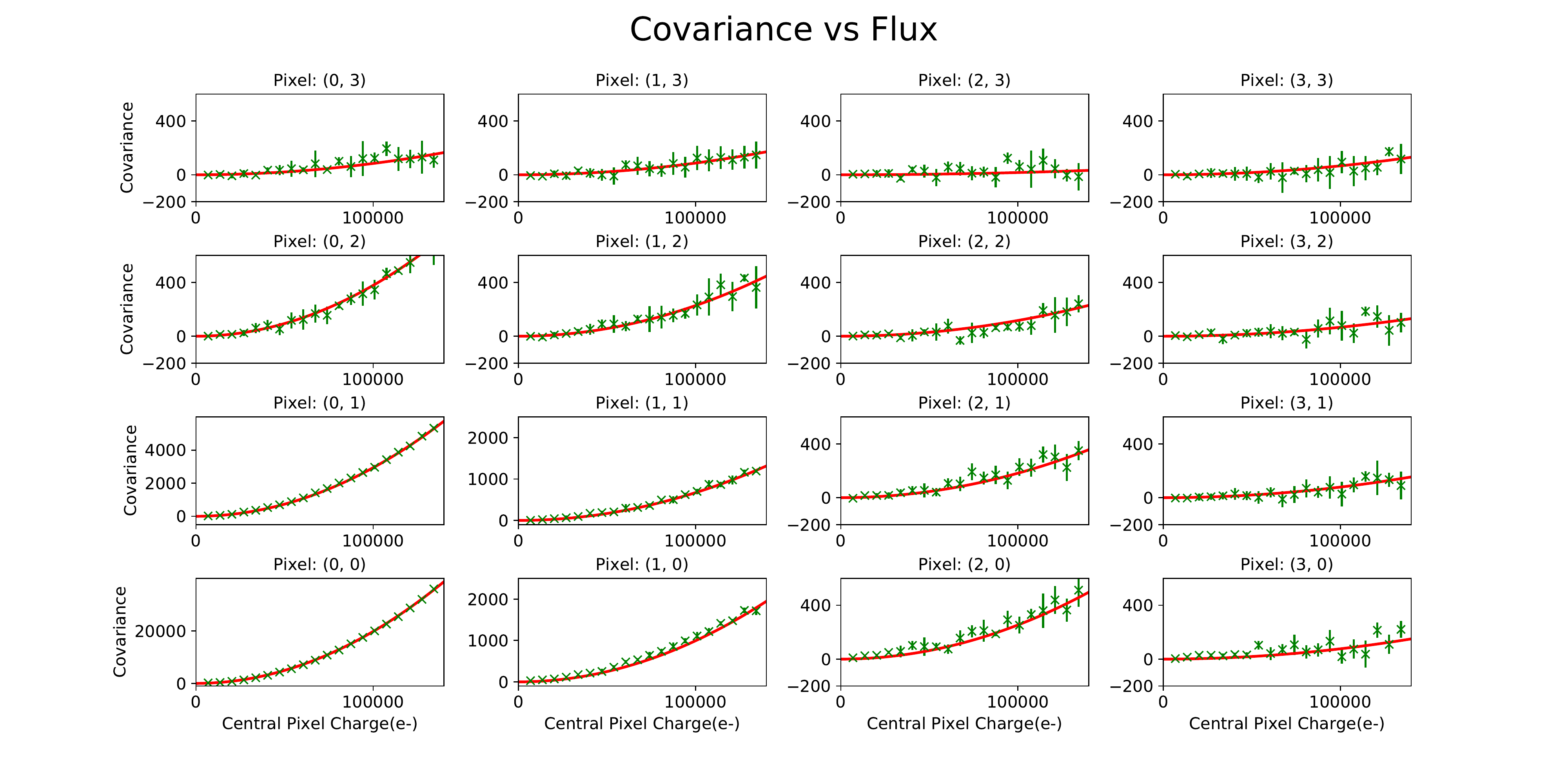}}
	\subfigure[b][Covariances vs Flux]{\includegraphics[trim=0.5in 0.0in 0.5in 0.0in,clip,width=0.99\textwidth]{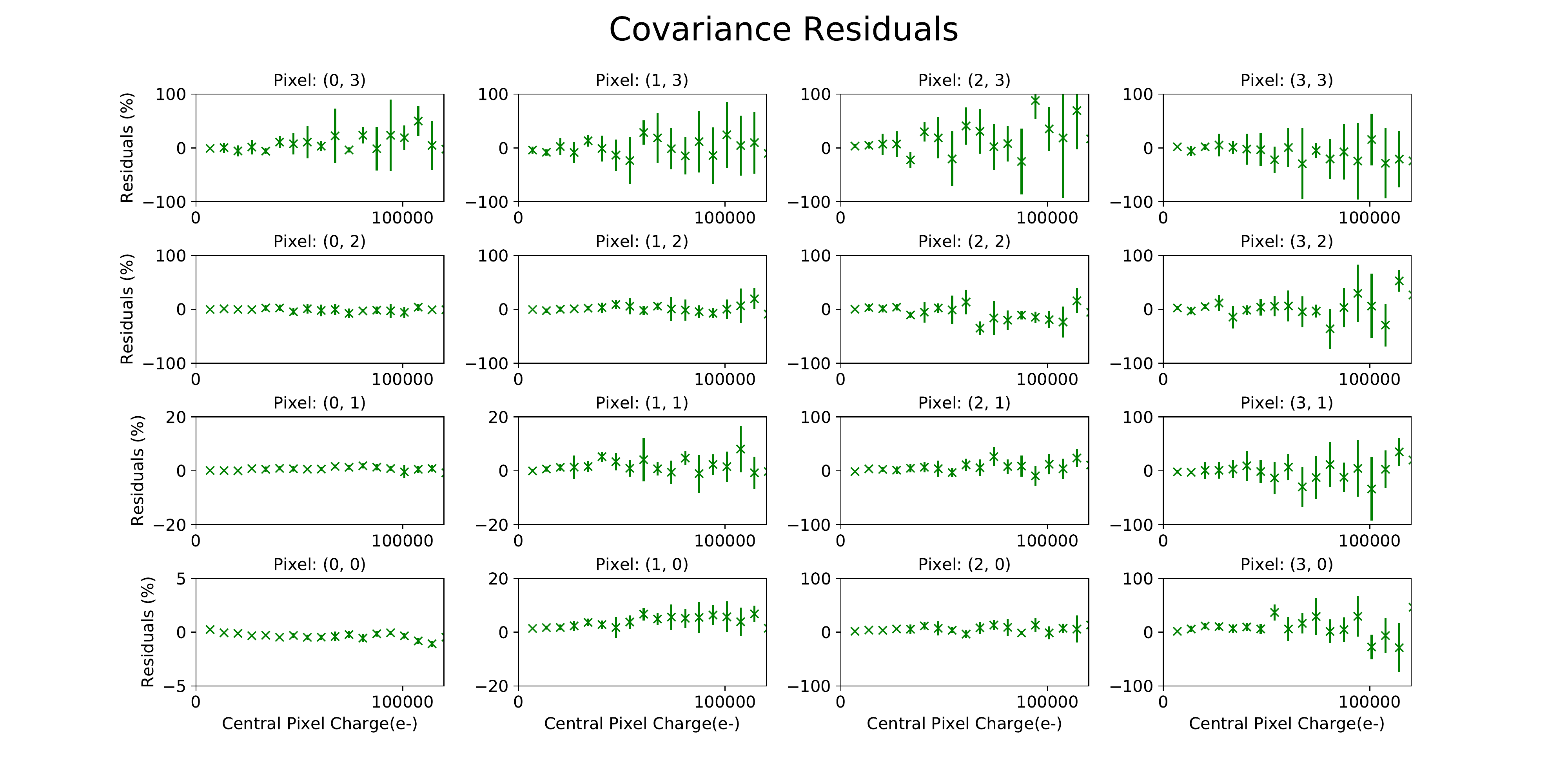}}
  \caption{Quadratic dependence of the covariances for the E2V detector. This is based on 100 flat pairs.  The red line is a quadratic fit - i.e. $\rm C_{ij} = \alpha f^2$, where the value of $\alpha$ is determined to minimize the sum of squared differences.  The residuals are the percentage difference between this quadratic fit and the data (green crosses).    The quadratic fit fits the data well, and there is no apparent systematic trend in the residuals indicating the need for higher order terms.  The residuals in the serial direction are smaller than in the case of the ITL detector, consistent with the smaller values of serial charge transfer inefficiency (CTI).}
  \label{Correlations_Quad_E2V}
  \end{figure}

\begin {figure}[p]
	\centering
	\subfigure[b][ITL - Covariances as measured]{\includegraphics[trim=0.0in 0.0in 0.0in 0.0in,clip,width=0.49\textwidth]{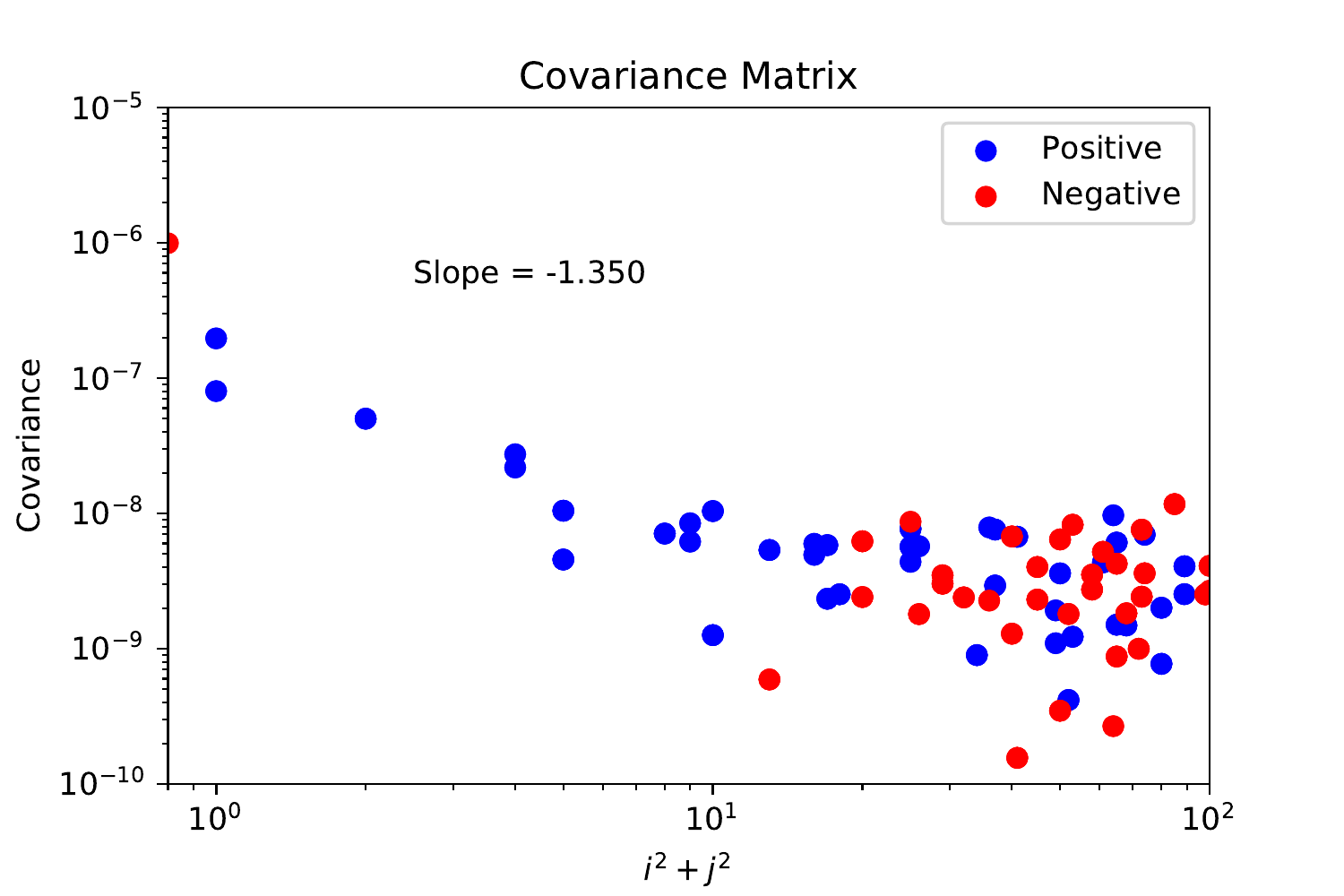}}
	\subfigure[b][ITL - Model used for pixels more than 3 away.]{\includegraphics[trim=0.0in 0.0in 0.0in 0.0in,clip,width=0.49\textwidth]{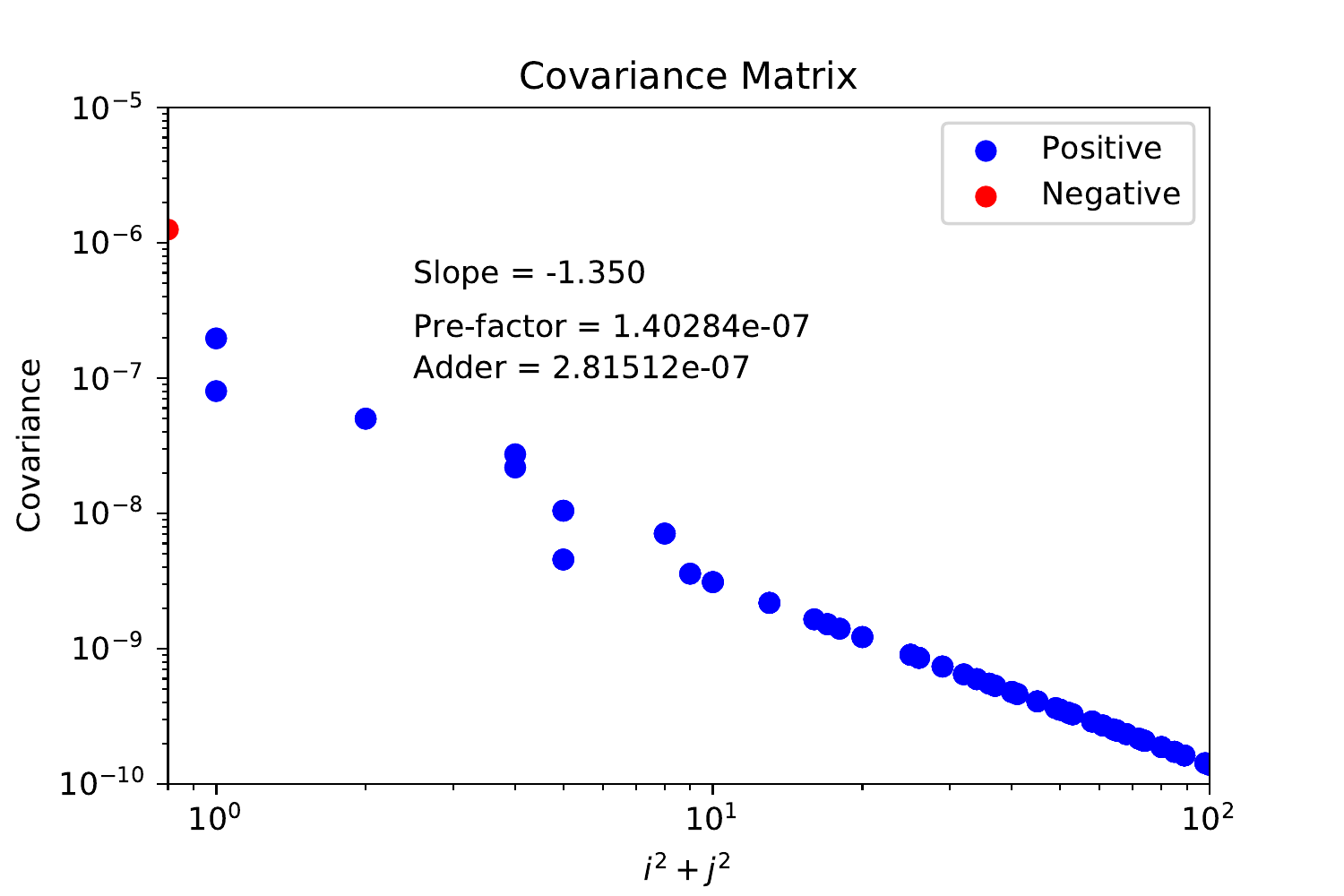}}
  \caption{Measured covariances for distant pixels are small and the measurements are noisy.  The 'Model' option uses a fit to the covariances for more distant pixels (in this case 3 or more pixels away from the origin).  Having a model also allows one to sum the model to infinity and include this value in the 'Zero' option.}
  \label{Model}
  \end{figure}

\begin{itemize}
  \item Gain: This involves correcting the aforementioned error in the gain calculation.  The code fits the PTC with a cubic curve and keeps only the linear part as the gain, as seen in Figure \ref{Gains}.
  \item Quad: Referring to Figure \ref{Correlations_Quad_E2V}, we can see that the covariances are a quadratic function of the flux.  The existing code takes each measurement of the covariance at a given flux, divides by the square of the flux, and then averages these measurements.  This option instead fits a quadratic curve through the measurements, as shown in Figures \ref{Correlations_Quad_ITL} and \ref{Correlations_Quad_E2V}.  Both methods work, but we find that the quadratic fit method gives somewhat better results.  
  \item Zero:  The covariance matrix should sum to zero, as lost variance from the central pixel is transferred to covariances in surrounding pixels.  However, for a number of reasons, it typically does not.  The $\rm C_{00}$ covariance value is basically the difference between the linear growth of the Poisson variance and the measured variance.  Therefore small errors in the gain can result in large errors in this term.  What is done in this option is to adjust the $\rm C_{00}$ covariance value to be equal to the negative of the sum of all of the other covariance terms, which forces the covariance matrix to sum to zero.  This is basically equivalent to binning the pixels together to create large pixels which are immune to the BF effect.  In the limit of this case, the central pixel variance becomes purely Poisson, and the surrounding covariances are zero.  This change is found to have the largest impact on improving the BF correction.
  \item Model: Measured covariances for distant pixels are small and the measurements are noisy.  This change uses a fit to the covariances for more distant pixels.  Having a model of the covariances as a function of distance from the central pixel also allows one to integrate the model to infinity and include this value in the sum of covariances in the above 'Zero' option, so this is done as well.  The covariance values that result from this option are shown in Figure \ref{Model}.
\end{itemize}

\begin {figure}[p]
	\centering
	\includegraphics[trim=2.0in 3.0in 1.5in 0.0in,clip,width=0.95\textwidth]{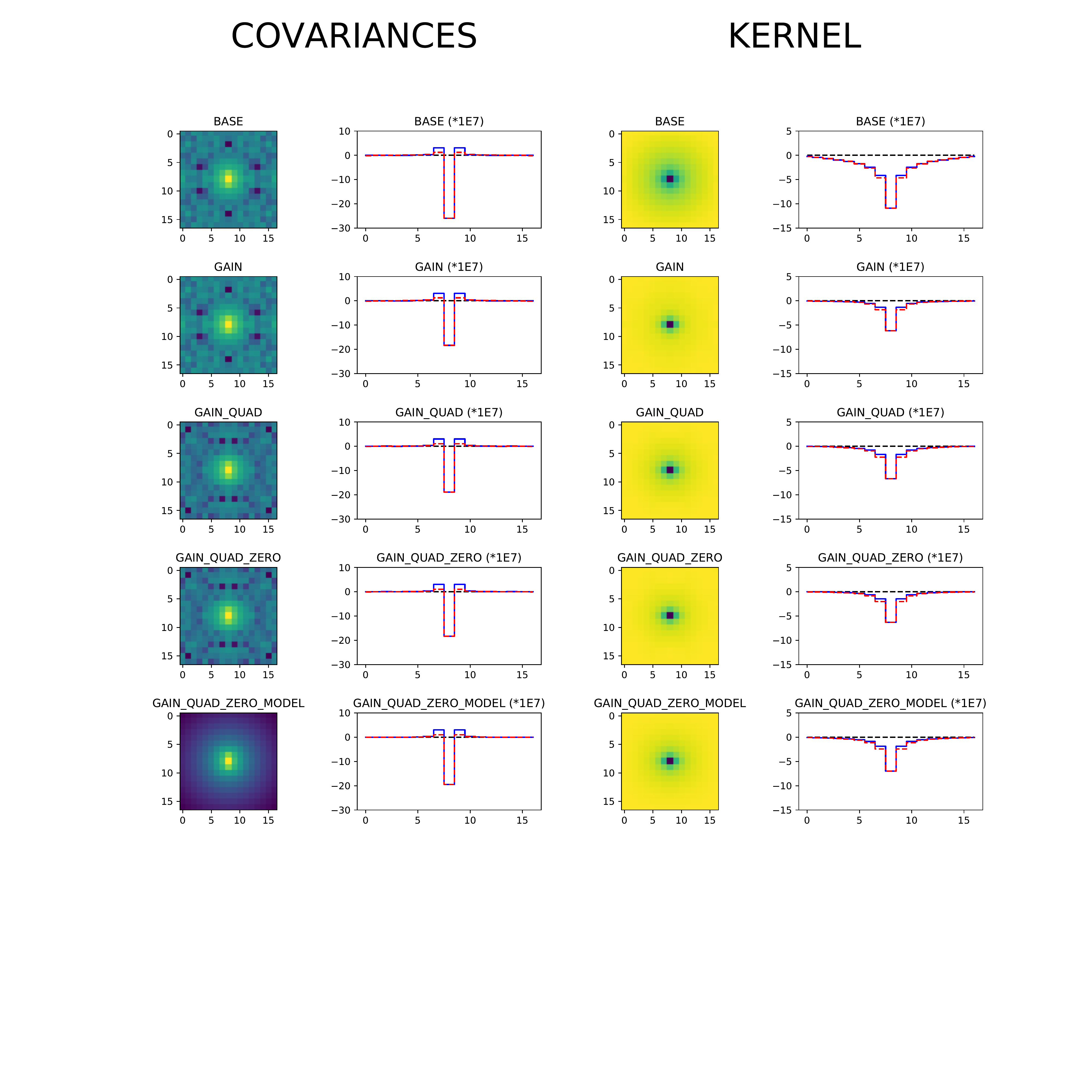}
  \caption{Covariances and resultant kernels with the different code changes applied sequentially to the E2V sensor. On the left are 2D plots and 1D slices of the covariance matrix, and on the right are 2D plots and 1D slices of the resultant kernels.  The central pixel is at (8,8) in these maps.  ``Baseline'' is the existing code.  ``Gain'' corrects the gain determination.  ``Quad'' uses the quadratic fit to the covariances instead of simple averaging.  ``Zero'' adjusts the $\rm C_{00}$ covariance term to force the covariance matrix to have zero sum.  ``Model'' uses a fit to the covariance terms for pixels which are 3 pixels or more away from the central pixel.  The baseline code results in a $\rm C_{00}$ value which is too negative, which is the cause of the overcorrection.}
  \label{Corr_Kernels}
  \end{figure}

Figure \ref{Corr_Kernels} shows the covariance matrices and kernels that result from these code changes.  The baseline code results in a $\rm C_{00}$ value which is too negative, which is the cause of the overcorrection.  The improved smoothness in the 'Model' option is apparent in the plot of the covariance matrix.

\begin {figure}[p]
	\centering
	\subfigure[b][ITL Gain Quad Zero Model]{\includegraphics[trim=1.0in 0.0in 1.5in 0.0in,clip,width=0.95\textwidth]{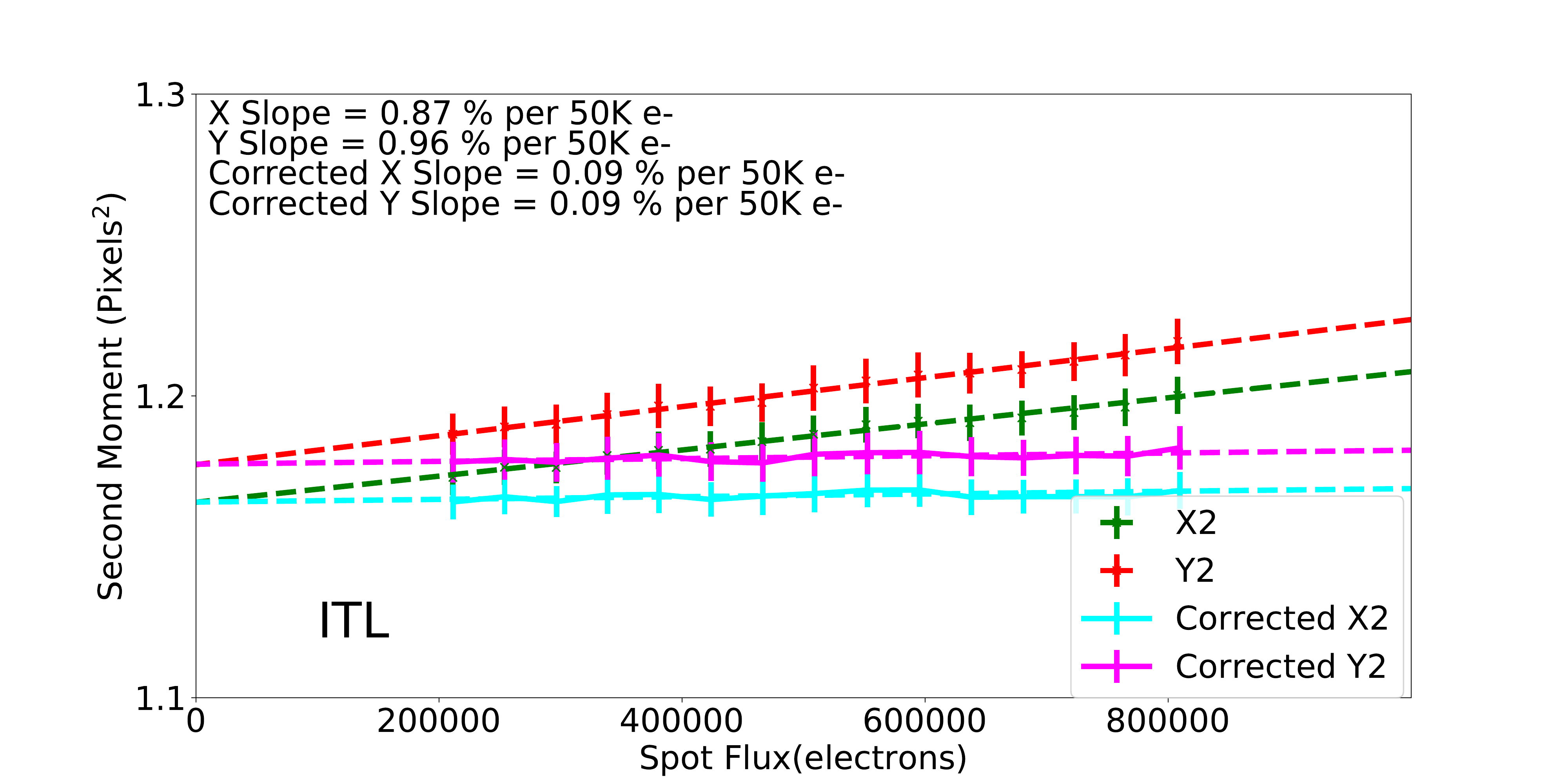}}
	\subfigure[b][E2V Gain Quad Zero Model]{\includegraphics[trim=1.0in 0.0in 1.5in 0.0in,clip,width=0.95\textwidth]{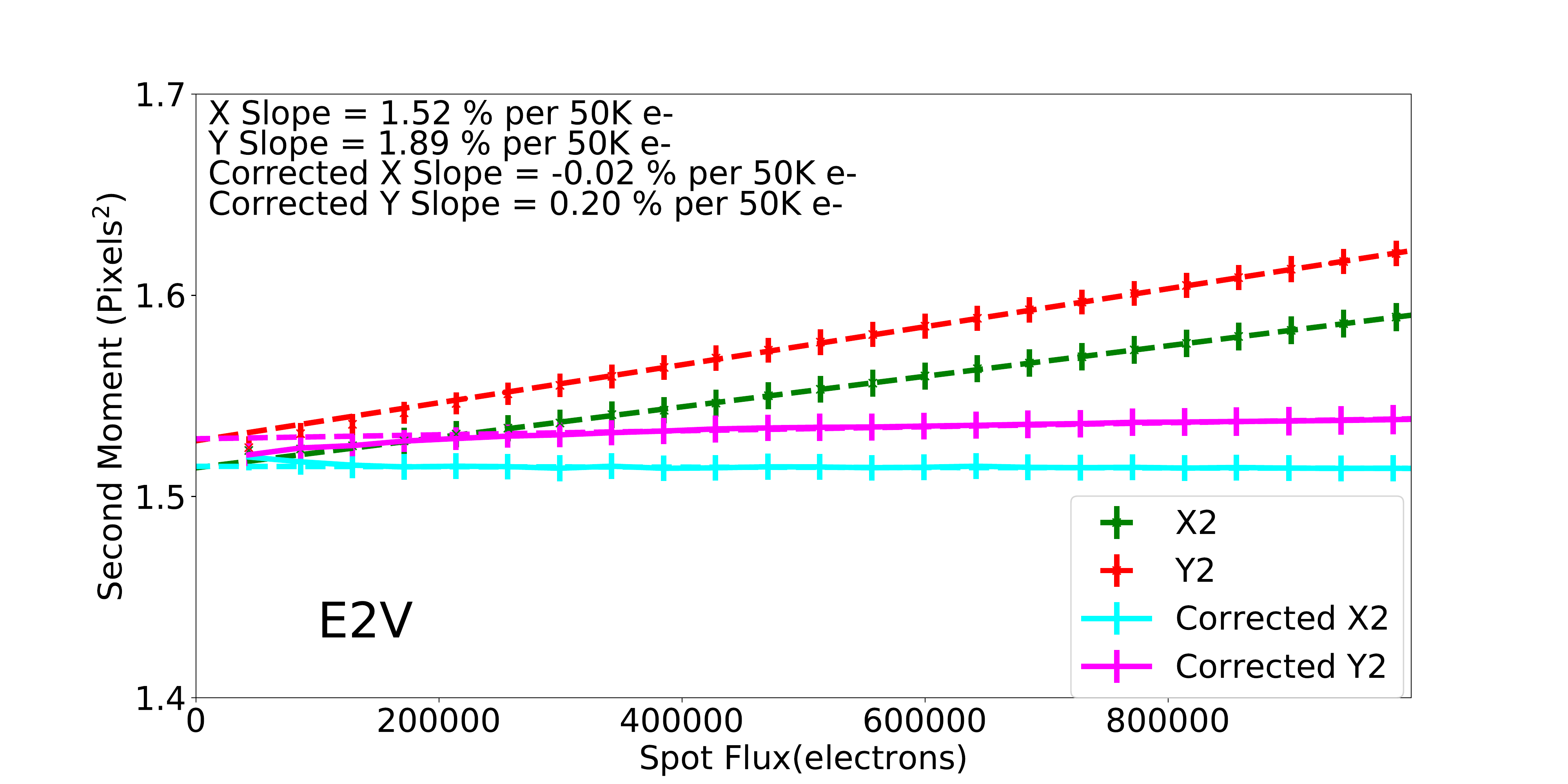}}

  \caption{Summary of BF correction results on the both detectors using the code improvements described in this work.  With the best improvements, we are correcting approximately 90\% of the effect.}
  \label{BF_Corrections_Both}
  \end{figure}

The result of adding these changes sequentially is shown in Figure \ref{BF_Corrections_E2V}, and Figure \ref{BF_Corrections_Both} shows the summary on both detectors.  When all of these changes are applied, we succeed in correcting the BF effect at the 90\% level or better.

\section{Discussion}
We show from both high-level electrostatic simulations of the LSST detectors, as well as direct measurements of flat covariances, that the BF effect is sufficiently linear that the Coulton correction algorithm may be adequate for BF effect correction.  We also show, assuming sufficient care is paid to calculating the correction kernel, that the Coulton algorithm does in fact correct 90\% or more of the BF effect on measured spots images.   How well do we need to correct the effect to meet the LSST science goals?  Mandelbaum \cite{Mandelbaum2015} estimates that an uncorrected BF effect will result in a multiplicative shear systematic error of approximately $\rm m = 0.06$.  Correcting 90\% of the effect should get us down near $\rm m=0.006$  To achieve the desired levels of $\rm m \approx 0.001 - 0.003$, we need to do a factor of 2-5 times better.   We believe that further improvements are possible.  The most urgent area for further improvement is clearly to remove the effects of CTI in the serial direction.  Algorithm improvement will continue as more data becomes available from a larger sample of sensors.

\section{Acknowledgements}
Andrew Bradshaw has been a key partner in building the UC Davis CCD lab, developing the software, and making the CCD measurements.  Kirk Gilmore's help in setting up the hardware and software for the CCDs from both vendors has also been invaluable.  Perry Gee has provided key support in software and networking.  Merlin Fisher-Levine has patiently helped with my understanding and bring-up of the LSST DM software stack.  Of course, Tony Tyson's vision in building the CCD lab and constant support are much appreciated.  Financial support from DOE grant DE-SC0009999 and Heising-Simons Foundation grant 2015-106 are gratefully acknowledged.

\bibliographystyle{unsrt}
\bibliography{ccd}

\begin{thebibliography}{10}

\bibitem{Coulton_2018}
William~R. Coulton, Robert Armstrong, Kendrick~M. Smith, Robert~H. Lupton, and
  David~N. Spergel.
\newblock {Exploring the Brighter-fatter Effect with the Hyper Suprime-Cam}.
\newblock {\em The Astronomical Journal}, 155(6):258, May 2018.

\bibitem{LSST_2019}
{\v Z}.~{Ivezi{\'c}}, S.~M. {Kahn}, J.~A. {Tyson}, B.~{Abel}, E.~{Acosta},
  R.~{Allsman}, D.~{Alonso}, Y.~{AlSayyad}, S.~F. {Anderson}, J.~{Andrew}, and
  et~al.
\newblock {LSST: From Science Drivers to Reference Design and Anticipated Data
  Products}.
\newblock {\em \apj}, 873:111, March 2019.

\bibitem{oconnor2019uniformity}
P.~O'Connor.
\newblock {Uniformity and Stability of the LSST Focal Plane}, 2019.
\newblock arXiv:1907.00995.

\bibitem{oconnor_2016}
P.~O'Connor, P.~Antilogus, P.~Doherty, J.~Haupt, S.~Herrmann, M.~Huffer,
  C.~Juramy-Giles, J.~Kuczewski, S.~Russo, C.~Stubbs, and R.~Van Berg.
\newblock {Integrated system tests of the LSST raft tower modules}.
\newblock In Andrew~D. Holland and James Beletic, editors, {\em High Energy,
  Optical, and Infrared Detectors for Astronomy VII}, volume 9915, pages 327 --
  338. International Society for Optics and Photonics, SPIE, 2016.

\bibitem{ITL_website}
University of~Arizona.
\newblock {Imaging Technology Laboratory}, 2016.
\newblock http://www.itl.arizona.edu.

\bibitem{E2V_website}
{Teledyne E2V}, 2019.
\newblock https://www.teledyne-e2v.com/products/space/.

\bibitem{antilogus2014}
P.~{Antilogus}, P.~{Astier}, P.~{Doherty}, A.~{Guyonnet}, and N.~{Regnault}.
\newblock {The brighter-fatter effect and pixel correlations in CCD sensors}.
\newblock {\em Journal of Instrumentation}, 9:C03048, March 2014.

\bibitem{guyonnet2015}
A.~{Guyonnet}, P.~{Astier}, P.~{Antilogus}, N.~{Regnault}, and P.~{Doherty}.
\newblock {Evidence for self-interaction of charge distribution in
  charge-coupled devices}.
\newblock {\em \aap}, 575:A41, March 2015.

\bibitem{gruen2015}
D.~{Gruen}, G.~M. {Bernstein}, M.~{Jarvis}, B.~{Rowe}, V.~{Vikram}, A.~A.
  {Plazas}, and S.~{Seitz}.
\newblock {Characterization and correction of charge-induced pixel shifts in
  DECam}.
\newblock {\em Journal of Instrumentation}, 10:C05032, May 2015.

\bibitem{rasmussen2016}
Andrew {Rasmussen}, Augustin {Guyonnet}, Craig {Lage}, Pierre {Antilogus},
  Pierre {Astier}, Peter {Doherty}, Kirk {Gilmore}, Ivan {Kotov}, Robert
  {Lupton}, Andrei {Nomerotski}, Paul {O'Connor}, Christopher {Stubbs}, Anthony
  {Tyson}, and Christopher {Walter}.
\newblock {High fidelity point-spread function retrieval in the presence of
  electrostatic, hysteretic pixel response}.
\newblock volume 9915 of {\em Society of Photo-Optical Instrumentation
  Engineers (SPIE) Conference Series}, page 99151A, Aug 2016.

\bibitem{astier2019}
Pierre {Astier}, Pierre {Antilogus}, Claire {Juramy}, R{\'e}my {Le Breton},
  Laurent {Le Guillou}, and Eduardo {Sepulveda}.
\newblock {The shape of the photon transfer curve of CCD sensors}.
\newblock {\em \aap}, 629:A36, Sep 2019.

\bibitem{Lage_2017}
C.~{Lage}, A.~{Bradshaw}, and J.~A. {Tyson}.
\newblock {Measurements and simulations of the brighter-fatter effect in CCD
  sensors}.
\newblock {\em Journal of Instrumentation}, 12:C03091, Mar 2017.

\bibitem{Poisson-CCD-paper}
C.~{Lage}.
\newblock {Poisson-CCD: A dedicated simulator for modeling CCDs}, 2019.
\newblock arXiv:1911.09038.

\bibitem{Poisson-CCD-code}
C.~{Lage}.
\newblock {Poisson solver for LSST CCDs}, November 2019.
\newblock https://github.com/craiglagegit/Poisson\_CCD.

\bibitem{CCD-Physical-Analysis}
C.~{Lage}.
\newblock {Physical and electrical analysis of LSST sensors}, 2019.
\newblock In preparation.

\bibitem{sze1981physics}
S.M. Sze.
\newblock {\em Physics of Semiconductor Devices}.
\newblock Wiley-Interscience publication. John Wiley \& Sons, 1981.

\bibitem{tyson2014}
J.~A. {Tyson}, J.~{Sasian}, K.~{Gilmore}, A.~{Bradshaw}, C.~{Claver},
  M.~{Klint}, G.~{Muller}, G.~{Poczulp}, and E.~{Resseguie}.
\newblock {LSST optical beam simulator}.
\newblock In {\em High Energy, Optical, and Infrared Detectors for Astronomy
  VI}, volume 9154 of {\em Society of Photo-Optical Instrumentation Engineers
  (SPIE) Conference Series}, page 915415, July 2014.

\bibitem{2018arXiv181203248B}
James {Bosch}, Yusra {AlSayyad}, Robert {Armstrong}, Eric {Bellm}, Hsin-Fang
  {Chiang}, Siegfried {Eggl}, Krzysztof {Findeisen}, Merlin {Fisher-Levine},
  Leanne~P. {Guy}, Augustin {Guyonnet}, {\v{Z}}eljko {Ivezi{\'c}}, Tim
  {Jenness}, G{\'a}bor {Kov{\'a}cs}, K.~Simon {Krughoff}, Robert~H. {Lupton},
  Nate~B. {Lust}, Lauren~A. {MacArthur}, Joshua {Meyers}, Fred {Moolekamp},
  Christopher~B. {Morrison}, Timothy~D. {Morton}, William {O'Mullane}, John~K.
  {Parejko}, Andr{\'e}s~A. {Plazas}, Paul~A. {Price}, Meredith~L. {Rawls},
  Sophie~L. {Reed}, Pim {Schellart}, Colin~T. {Slater}, Ian {Sullivan},
  John.~D. {Swinbank}, Dan {Taranu}, Christopher~Z. {Waters}, and W.~M.
  {Wood-Vasey}.
\newblock {An Overview of the LSST Image Processing Pipelines}.
\newblock {\em arXiv e-prints}, page arXiv:1812.03248, Dec 2018.

\bibitem{Mandelbaum2015}
R.~{Mandelbaum}.
\newblock {Instrumental systematics and weak gravitational lensing}.
\newblock {\em Journal of Instrumentation}, 10(5):C05017, May 2015.

\end{thebibliography}

\end{document}